RESEARCH ARTICLE

# Global and country-specific mainstreaminess measures: Definitions, analysis, and usage for improving personalized music recommendation systems

Christine Bauer⊙*, Markus Schedl⊙

Institute of Computational Perception, Johannes Kepler University Linz, Linz, Austria

⊙ These authors contributed equally to this work.
* christine.bauer@jku.at

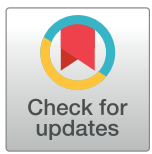

**Data Availability Statement:** We use the publicly available LFM-1b dataset of user-generated listening events from Last.fm. It can be downloaded from a dedicated web page: http://www.cp.jku.at/datasets/LFM-1b Schedl, Markus (2016). The LFM-1b Dataset for Music Retrieval and Recommendation. ACM International Conference on Multimedia Retrieval (ICMR), 103-110.

**Funding:** This work was supported by Austrian Science Fund (FWF): V579, https://www.fwf.ac.at,

## Abstract

### Relevance

Popularity-based approaches are widely adopted in music recommendation systems, both in industry and research. These approaches recommend to the target user what is currently popular among all users of the system. However, as the popularity distribution of music items typically is a long-tail distribution, popularity-based approaches to music recommendation fall short in satisfying listeners that have specialized music preferences far away from the global music mainstream. Addressing this gap, the contribution of this article is three-fold.

### Definition of mainstreaminess measures

First, we provide several quantitative measures describing the proximity of a user's music preference to the music mainstream. Assuming that there is a difference between the *global* music mainstream and a *country-specific* one, we define the measures at two levels: relating a listener's music preferences to the global music preferences of all users, or relating them to music preferences of the user's country. To quantify such music preferences, we define a music item's popularity in terms of *artist playcounts* (APC) and *artist listener counts* (ALC). Moreover, we adopt a *distribution-based* and a *rank-based* approach as means to decrease bias towards the head of the long-tail distribution. This eventually results in a framework of 6 measures to quantify music mainstream.

### Differences between countries with respect to music mainstream

Second, we perform in-depth quantitative and qualitative studies of music mainstream in that we (i) analyze differences between countries in terms of their level of mainstreaminess, (ii) uncover both positive and negative outliers (substantially higher and lower country-specific popularity, respectively, compared to the global mainstream), analyzing these with a mixed-methods approach, and (iii) investigate differences between countries in terms of listening preferences related to popular music artists. We conduct our studies and experiments using the standardized LFM-1b dataset, from which we analyze about 800,000,000 listening events shared by about 53,000 users (from 47 countries) of the music streaming platform Last.fm. We show that there are substantial country-specific differences in listeners' music consumption behavior with respect to the most popular artists listened to.

### Rating prediction experiments

Third, we demonstrate the applicability of our study results to improve music recommendation systems. To this end, we conduct rating prediction experiments in which we tailor recommendations to a user's level of preference for the music mainstream using the proposed 6 mainstreaminess measures: defined by a distribution-based or rank-based approach, defined on a global level or on a country level (for the user's country), and for APC or ALC. Our approach roughly equals a hybrid recommendation approach in which a demographic filtering strategy is implemented before collaborative filtering is performed. Results suggest that, in terms of rating prediction accuracy, each of the presented mainstreaminess definitions has its merits.

## Introduction

Nowadays, user-generated content is abundantly available online and the amount of available content increases tremendously on a daily basis [1]. Such consumable content is versatile and includes, for instance, news, videos, music, and photographs. The opportunity to access a large amount of content also has its downsides, because it frequently leads to information [2] or choice overload [3] and people do not know what to choose or do not find the content that they are interested in. Thus, recommender systems (also known as recommendation systems) have become important tools because they assists users in searching, sorting, and filtering the massive amount of any kind of online content [4]. As a result, they help decreasing the information and choice overload problems, too. Recommender systems play an important role in people's everyday life and support versatile activities such as shopping [5–7] or consuming news [1], movies [8, 9], and music [10–12].

For example, for the longest time, access to music recordings was restricted to local availability of their physical representations (e.g., vinyl, tape, CD). Usually, there was only a certain amount of recordings available at home. Now, using online music platforms such as YouTube, Spotify, Pandora, or iTunes, users have access to tens of millions of music recordings [13]. Music recommender systems (MRS) have been adopted by industry to assist listeners in navigating the catalogs of available music recordings and to serve them with suggestions of items that may fit the respective user's preferences [14]. Today, MRS have become a significant research area [10, 15]. Also music industry establishes research teams dedicated to the topic, because MRS are important drivers for their businesses [14].

Yet, from the user's perspective, current MRS frequently produce unsatisfactory recommendations [16]. An ideal MRS proposes "the right music, to the right user, at the right moment" [17]. However, this is a complex task because a multitude of factors influence a user's music preferences in a given situation [16, 18]. The relationships between music preferences and, for instances, demographics [19–22], personality traits [23–25], social influences [26–28], or listening habits [29, 30] have been investigated. Besides such person-related characteristics, situation-related factors influence a user's music preferences. Examples are user activity [31, 32], temporal aspects [33], or weather [34]. MRS that consider situation-related factors are typically referred to as "context-aware MRS" because systems that consider the context that they are used in and/or their users' context and adapt the system's operations to the current context without explicit user intervention are referred to as "context-aware systems" [35]. Context is "any information that can be used to characterize the situation of an entity" [36]. Thereby, context information may be derived from various sources (e.g., from a user's personal information on social media or from sensors such as accelerometer, gyroscope, or noise meter) including a user's active input [37]. A few years ago, context-awareness began to receive considerable attention in research on MRS (e.g., [38, 39]). Still, the multitude of relevant factors that influence music preferences are not considered in sufficient depth in current MRS [16]. Besides a few music players where a user can specify his or her mood or activity to tailor the music recommendations accordingly, to the best of our knowledge, no fully automated context-aware MRS has been released to the public yet [15]. Instead, most MRS rely mainly on the concept of user-item interactions (in collaborative filtering) or on information about the music items (in content-based filtering) [15, 40–43]. Both, in research and practice, most MRS largely disregard the variety of contextual factors influencing a user's music preferences [16].

The widely disregarded context in MRS research that we focus on in this work are the various country-specific mechanisms that affect a user's music preferences and consumption behavior. For instance, music preferences are shaped by cultural aspects [44, 45] and cultures vary across countries [46]. Even music perceptions vary across cultures [47–50]. Furthermore, national market structures are different across countries, which includes, for instance, distribution channels as well as legislation, subsidizing, advertising campaigns, local radio airplay, quotas for national artists on radio [51–53]. For example, in the European countries, preferences for pop music diverge rather than converge [45]. In other words, country-specific aspects strongly shape users' music preferences and music consumption behavior.

Against this background, one could expect that music artist's popularity differs across countries. In fact, though, the most popular artists are almost the same across countries. This phenomenon is related to the long-tail distribution of popularity. There is a strong concentration of globally highly popular artists (the short head of the distribution), but also a long tail of comparatively far less popular artists (known as the "long tail phenomenon" [54, 55]). At the same time, there are music artists that are very popular in one country but do not reach this high-popularity status on global scale; from the global perspective, these artist are not very popular and are to be found in the long tail of the popularity distribution [18, 56]. For instance, among Finnish listeners, users who like the artists Stam1na, Kotiteollisuus, or Katariina Hänninen are national top hit consumers, whereas, from a global perspective, these artists are much less popular and constitute a niche [56].

For MRS, this has the following implications: The long tail phenomenon implicates that it is more probable that a random user likes a very popular music item than one of the far less popular items [10, 57]. As a result, popularity-based recommendation approaches are widely adopted by MRS in particular to complement other approaches especially for cold start situations (i.e., for users new to the MRS under consideration [32, 58]). While such popularity-based recommendations may be successful when targeting listeners inclined to music items that are generally very popular (i.e., the "music mainstream"), they are not satisfying when addressing people with specialized music preferences that are not considered mainstream [16]. Most previous MRS approaches that exploit item popularity and user mainstreaminess [29, 59] measure a user's preference for mainstream music by dividing the respective user's consumption of each music item by its overall consumption by all users. Using such a fraction-based approach disproportionately privileges the music items that constitute the absolute top hits in the dataset (the head). As a result, the approach does not work well for listeners less inclined to mainstream music and, thus, leads to MRS with lower performance (in terms of rating prediction accuracy) for niche consumers. In earlier work [60], we proposed a distribution-based and a rank-based approach to measure a user's preference for mainstream music in order to counteract the bias towards the very top items in the popularity distribution. In further work [18, 56], we demonstrated that replacing the global perspective of music popularity (i.e., considering popularity among all users globally) by a country-specific perspective (i.e., considering popularity among the users of a specific country) is a viable approach for providing recommendations that better satisfy a wide variety of users, i.e., users covering a wider range of music preferences (not only global top hit consumers).

In the present work, we build upon the framework presented in [56, 60], streamline, and extend it. More precisely, we use the distribution- and rank-based approaches for measuring a user's preference for mainstream music (his/her *mainstreaminess*) that proved promising in earlier work [56, 60] and improve their definitions, for instance, by normalizing for better comparison of the approaches. In addition, besides *artist playcounts* (APC), we introduce *artist listener counts* (ALC) as a measure for music popularity. We further define a global and a country-specific version for a user's mainstreaminess. In total this leads to 6 variants of mainstreaminess measures that we analyze in this work (distribution- vs. rank-based, APC vs. ALC, and global vs. country perspective).

In addition, using these measures in a rating prediction approach for evaluation, we assess the performance of a recommender system by letting the system predict ratings (or consumption levels) for unknown user-item pairs and measuring its prediction error [61]. We investigate the performance differences (in terms of rating prediction accuracy) realized for users when both their country and their global or country-specific mainstream reflection in listening behavior are considered. To this end, we use a subset of the LFM-1b dataset [13] of user-generated listening events (53,259 Last.fm users from 47 countries). Country is considered a proxy for national culture in the present study. Being aware that the concept of national culture has been criticized for equating culture with nation and leaving aside ethnic aspects [62, 63], we emphasize that next to cultural aspects also national market structures contribute to users' music consumption preferences and behavior. Thus, country as proxy seems reasonable for the study at hand.

This article delivers three main contributions:

1. We provide several quantitative measures of the user's mainstreaminess, i.e., the closeness of a user's music preference to the music mainstream. Assuming that there is a difference between a global mainstream and a country-specific one, we quantify mainstreaminess on a global and a country level. We base our definitions of mainstreaminess on (i) similarities of popularity distributions and (ii) rank-order correlations between profiles of user, country, and global preference.

2. We conduct in-depth quantitative as well as qualitative studies of music mainstream as evidenced in user-generated listening data from Last.fm. We uncover considerable country-specific differences in listening behavior and preference. Comparing these country-specific differences to the global mainstream, we show that (i) in some countries, users' music consumption behavior corresponds to the global one, resulting in an overlap of country-specific and global mainstream, (ii) in some countries, besides the global mainstream an additional country-specific mainstream has developed, and (iii) some countries do not show a clear picture concerning music mainstream consumption behavior and artist popularity. Furthermore, we identify and discuss both positive and negative artist outliers (substantially higher and lower country-specific popularity, respectively) for selected countries.

3. We demonstrate how considering a user's mainstreaminess level and country—which we use as proxy for cultural background (here: national culture)—in the personalized music recommendation process can notably improve accuracy of rating prediction, compared to a one-fits-all solution without country information.

The remainder of this article is structured as follows: We start with an outline of the conceptual foundations of our work. Then, we detail the methods and procedures employed in the research at hand. Next, we report and discuss the results of our studies. In the final section, we conclude with a summary and an outlook to future work.

## Conceptual foundations

This research foremost connects to the area of music recommendation systems as well as to the aspects of popularity and mainstream in the music domain, and their consideration in recommendation systems. We provide a conceptual foundation and discuss related work about the former in Section *Music recommendation systems*, about the latter in Section *Music popularity and mainstreaminess* and Section *Popularity and mainstreaminess in music recommendation*, respectively.

### Music recommendation systems

Recommender systems (also known as recommendation systems) are computer systems which, based on their users' preferences, provide suggestions for items deemed interesting to the target user, assisting him or her in various decision-making processes (e.g., relating to what products to buy, what music to listen to, or what online news to read) [61]. The general term used to denote what the system recommends to users is "item" [61]. Examples for such items include hotels for an upcoming vacation, friends in an online social network, exercises on an e-learning platform, or books to read. Three main components are vital to build and maintain a recommender system: users, items, and algorithms to match the former two.

In a music recommendation system (MRS), users are most commonly the music listeners and items are the music entities that can be recommended, for instance, performers, albums, or individual music pieces [14]. Note that there also exist MRS that support music creators, e.g., to recommend music building blocks such as drum loops (e.g., [64]). Nevertheless, we focus on the music consumers in this work. The user-item matching algorithms in recommender systems are typically based on one of the following approaches: content-based filtering (CBF), collaborative filtering (CF), or hybrid approaches that combine techniques from CBF and CF [65]. CBF approaches to MRS exploit item content descriptors, for instance, rhythm, tempo, instrumentation, lyrics, genre, or style of a music piece [15, 40–43], to build a user profile. Such descriptors are calculated or inferred from either the audio signal (using audio analysis techniques) [10, 66, 67], editorial metadata (e.g., genre or release year) [40, 68], user-generated content (e.g., tags or reviews) [69, 70], or annotations gathered via web content mining [71, 72]. Users are then modeled in terms of the music content they prefer and items similar according to these content descriptors are recommended.

In contrast, CF approaches to MRS do not rely on exogenous information about items or users, but on user-item interactions. Such interactions are interpreted as explicit or implicit feedback about users' preferences. The most common explicit feedback type is user ratings (e.g., provided on a 5-point Likert rating scale). Implicit preference feedback may be derived from listening patterns, e.g., number of listening events [13] or song skipping behavior [73]. CF approaches maintain a user profile for each user, which encodes this preference information. To recommend music items to a target user, his or her most similar users based on their user profile are identified and items listened to by them, but not known by the target user are suggested by the system [61]. This method is known as user-based CF. Item-based CF, in contrast, recommends items similar to the ones liked by the target user where similarity is computed over user ratings. An alternative, typically better performing method is model-based CF, which describes both users and items in a vector space whose dimensionality is lower than the number of users and the number of items [74]. Therefore, the computation of matches between items and users can be performed efficiently. Model-based CF is, hence, the de-facto standard in today's commercial recommender systems [9].

Hybrid approaches which commonly integrate CBF and CF techniques, or extend them with context information, aim at combining the advantages of the individual approaches, circumventing their shortcomings [10, 16]. In the case of CF, these limitations include the cold start problem (the system has no information about new users and new items) and data sparsity (users commonly rate only very few items in comparison to the size of the catalog from which the recommender system can suggest items, which typically range in the tens of millions of songs). Limitations of CBF include the hubness problem [75], which is caused by the high dimensionality of content feature descriptors, i.e., some items occur among the most similar items of a large number of other items in the catalog without actually being similar. Also, content-based techniques commonly perform inferior to CF in terms of accuracy [76].

### Music popularity and mainstreaminess

The popularity of items, in general, can be modeled as a distribution on a popularity curve [10], such as the one depicted in Fig 1, where the y-axis shows each item's value for popularity or demand, and items are sorted in decreasing order of popularity along the x-axis. The most popular items form the "head" of the distribution, whereas the least popular are referred to as "tail". The existence of a highly uneven distribution of demand for the most popular and the least popular items in an economy is referred to as the "long-tail economy" [54, 55]. Such a long-tail distribution of popularity is also typical for online music platforms, as could be demonstrated by [10]. The most popular music items are commonly dubbed "hits" [10], the "short head" [77], or "mainstream music" [10, 66, 78]—with all these terms referring to the general concept of popularity concentration. (Note: According to the Oxford Dictionaries, the term

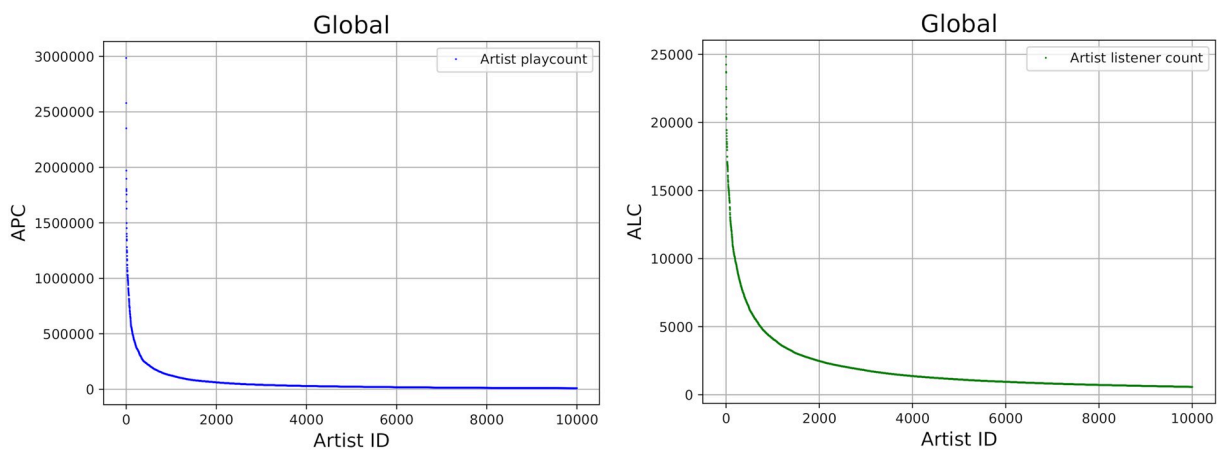

**Fig 1. Artist playcounts (APC) and artist listener counts (ALC) for the global top 10,000 artists.** Artist IDs (x-axis) are sorted by popularity values.

https://doi.org/10.1371/journal.pone.0217389.g001

*mainstream* is defined as "The ideas, attitudes, or activities that are shared by most people and regarded as normal or conventional").

There exist several ways to define and to measure popularity in the music domain. The most common ones are radio air plays, sales figures, downloads, or media coverage. In music streaming services, a song's, album's, or artist's popularity is often described in terms of total *playcount* of the item. This equals the number of listening events the item realizes in total by all listeners, cf. [10]. Likewise, the *listener frequency* (also known as *listener count*) can be used as a proxy for item popularity [79]. This equals the number of unique listeners of the item, regardless of their listening frequency. The listener count indicates the number of users that an artist reaches. For instance, an artist could have a high playcount and a low listener count if the high number of playcounts is realized by a limited number of listeners who listen to the same artists very frequently.

On the level of listeners, their inclination towards popular music can be described "in terms of the degree to which they prefer music items that are currently popular or rather ignore such trends" [11]. This inclination can also be referred to as "music mainstreaminess of a user" [11, 29]. It quantifies how strongly the target user's playcount of the music items under consideration (artists, albums, etc.) is in line with the respective playcount of the population at large.

## Popularity and mainstreaminess in music recommendation

Assuming that a random user is more likely to prefer a popular item of the short head of the popularity distribution than a less popular item of the long tail, popularity-based recommendation algorithms that recommend the overall most popular items are widely adopted [10, 57], especially in the absence of detailed user profiles (cold start). Besides music recommendation [32, 58], such algorithms are also common in product recommendation [5] or news recommendation [80].

As an alternative to considering only the most popular items for recommendation, popularity information about items can also be used to describe the mainstreaminess of users, which in turn allows to adjust the recommendations to the user's desired level of mainstreaminess. In fact, recent studies [29, 59] revealed that computing music mainstreaminess of users and integrating the resulting information into a CF recommender can yield better accuracy and rating prediction error than pure CF. However, this early work on user mainstreaminess for MRS is restricted to the use of fraction-based mainstreaminess measures (see Section *Introduction*). Such measures quantify user mainstreaminess as fractions of the target user's playcounts in relation to the playcounts of the entire user base, which disproportionately privileges the very top (head) items in the popularity distribution. As a result, adopting these fraction-based measures to model user mainstreaminess in CF approaches performs inferior to using latest mainstreaminess measurement approaches [60]. The most recent approaches calculate user mainstreaminess based on similarities between distributions of playcounts (between the target user and the overall population) or based on rank-order correlations [60]. In the latter case, the target user's as well as the entire population's playcount values are first converted into ranks. Subsequently, Spearman's rank-order correlation [81] between the resulting two rank vectors (user and global) yields the user's mainstreaminess score. Due to the superior performance of MRS that integrate user models based on these mainstreaminess measurement approaches, in our experiments we will extend these state-of-the-art approaches, building upon the framework presented in [56, 60], as mentioned in Section *Introduction* and detailed in the next section.

## Materials and methods

In the work at hand, we follow a three-step approach. First, we provide several *measures to quantify the mainstreaminess* in a user's or country's music listening behavior, in relation to both a global and a country-specific mainstream. We extend our previous work [56, 60] in that we streamline our mainstreaminess framework, and define and investigate distribution- and rank-based mainstreaminess measures on artist playcounts (APC) as well as on artist listener counts (ALC). We further normalize the distribution-based approaches for better comparison.

Second, we *analyze country-specific differences of the distribution of (mainstream) music*. We use the publicly available LFM-1b dataset of user-generated listening events from Last.fm [13] for this investigation. In particular, we (i) investigate differences between countries in terms of listening preferences related to popular music artists, (ii) analyze differences in mainstream preferences between countries, and (iii) uncover both positive and negative outliers (substantially higher and lower country-specific popularity, respectively) for selected countries and analyze the results with a qualitative approach.

Third, in line with common recommender systems evaluation, we perform *rating prediction experiments* using again the LFM-1b dataset. In particular, we analyze the performance of a state-of-the-art collaborative filtering recommender when *tailoring the recommendations to user groups* defined according to their level of *mainstreaminess* and their *cultural background*, for which we use country as proxy. More specifically, we use the introduced quantitative measures to gauge the mainstreaminess, both on a global level and on a country level. For both scopes, we group users according to their mainstreaminess levels into three (almost) equally sized classes (according to low, medium, and high mainstreaminess tertiles).

In the following, Section *Dataset* describes the sample of the LFM-1b dataset we use in our studies. In Section *Approaches to measure music mainstreaminess*, we detail the deployed approaches for mainstreaminess measurement. In Section *Approaches to analyze country-specific differences in music mainstream*, we introduce our approaches to investigate country differences in terms of mainstreaminess level and popular artists as well as to detect outliers. Section *Setup of rating prediction experiments* details the recommendation setup for the rating prediction experiments and outlines the evaluation metrics that we use to assess the quality of recommendations.

## Dataset

We use the LFM-1b dataset [13] that comprises 1,088,161,692 listening events of 120,322 unique users. It can be downloaded from a dedicated web page (http://www.cp.jku.at/datasets/LFM-1b). The essential part for our analyses is the user-artist-playcount matrix (UAM) containing the listening events of 120,175 unique users to 585,095 unique artists. The distribution of artist playcounts resembles a typical long-tail distribution [10]. Since our investigation focuses on country-specific differences, we consider a subset of the LFM-1b dataset, which only includes listening events of users who provided country information. To reduce the likelihood of less significant results due to a sample bias of users within a given country, we furthermore only consider countries with at least 100 users. The respective filtering of the dataset results in 53,259 users from 47 countries, who created a total of about 800 million listening events. In order to perform the evaluation of recommender systems via rating prediction (Section *Setup of rating prediction experiments*), we subsequently normalize the playcount values in the UAM to the range [0, 1], for each user individually.

Even though we are sure that the used LFM-1b dataset is highly valuable and appropriate for our analysis, we do not want to conceal its limitations. The LFM-1b dataset may not necessarily generalize to the population at large. For instance, the usage of Last.fm data introduces a community bias [82]. In terms of age distribution, it may also not be representative to the global population at large [83]. Similarly, the dataset is known to be biased with respect to gender (with a high percentage of male users in the sample). Listeners of classical music tend to be underrepresented [84], whereas listeners of the genres metal and alternative tend to be overrepresented on the Last.fm platform [85]. Still, the LFM-1b dataset provides an indication of the user composition of a typical online music platform and the proposed MRS approaches are meant to be used on such online music platforms. Accordingly, we consider the LFM-1b dataset an adequate source for our analyses.

**Approaches to measure music mainstreaminess**

When describing how well a user's listening preferences reflect those of an overall population, e.g., globally or within a country, what is considered *mainstream* depends on the selection of a population; this is a phenomenon which we will also show in our analysis. Consequently, we propose several quantitative measures for a user's music mainstreaminess, both on a global and on a country-specific level, depending on the selection of the population against which the user is compared. Our mainstreaminess measurement approaches are based on, and extend, our previous research [18, 56] in that we improve the formal framework and further consider not only *artist playcounts* (APC), but also *artist listener counts* (ALC) in the definitions of mainstreaminess. $APC_a$ of artist $a$ is defined as the sum of all single listening events to tracks by artist $a$. $ALC_a$ of an artist $a$ is defined as the number of unique listeners who listened to artist $a$ at least once. Our motivation to investigate APC and ALC as artist popularity measures is that both reflect a different kind of listening preference. While APC is an accumulative measure (more listening events, even by the same user, always lead to a higher value), ALC approximates the spread or overall familiarity of an artist (only more listeners contribute to a higher value). To illustrate this, Table 1 shows the top 30 artists in the entire LFM-1b dataset, once in terms of APC (left), once in terms of ALC (right). Artists that are familiar to a wide range of users (high ALC) do not necessarily accumulate a similarly high number of individual playcounts (APC). Prominent examples are Michael Jackson, The Doors, Johnny Cash, and U2. In contrast, artists such as Metallica, Lana Del Rey, System of a Down, Iron Maiden, or Depeche Mode accumulate a very high number of APCs from their fans, but are listened to by a lower variety of users (low ALC rank in comparison to APC rank).

Table 2 provides the formal definitions alongside the denominations/abbreviations of the mainstreaminess measures under investigation. We distinguish two kinds of measures: distribution-based and rank-based measures, each at two different scopes (global and country), using either APC or ALC as popularity measure. Note that computing the mainstreaminess using the combination of the rank-based approach and the ALC measure is impossible (see below). In total, we therefore define 6 measures.

The distribution-based measures rely on the symmetrized Kullback-Leibler divergence [86] (also known as relative entropy) between distributions of artist popularities; the rank-based measures are based on rank-order correlation according to Kendall's $\tau$ [87, 88]. The adoption of Kullback-Leibler divergence is motivated by the fact that it is a well-established method to compare (in our case discrete) distributions [89, 90]. We employ rank-order correlation because conversion of feature values to ranks has already been proven successful for music similarity tasks [91]. Please note that in contrast to our previous work [56, 60], we normalize the individual Kullback-Leibler divergences before symmetrizing them in order to constrain results to the well-defined range of [0, 1]. This facilitates comparison and results in distribution-based measures being represented as *distance functions*. Accordingly, a value of 0 means that there is no overlap between the compared artist popularity distributions; a value of 1

**Table 1. Top 30 artists in the entire LFM-1b dataset in terms of artist playcount (APC) and artist listener count (ALC).**

| Artist | APC | Artist | APC |
|---|---|---|---|
| The Beatles | 3,838,604 | Coldplay | 48,640 |
| Radiohead | 3,437,326 | Radiohead | 44,707 |
| Pink Floyd | 2,990,318 | Daft Punk | 44,356 |
| Coldplay | 2,576,390 | Nirvana | 41,488 |
| Daft Punk | 2,523,537 | Red Hot Chili Peppers | 40,421 |
| Muse | 2,460,597 | Queen | 39,874 |
| Metallica | 2,401,945 | Muse | 39,726 |
| Arctic Monkeys | 2,345,951 | The Rolling Stones | 39,033 |
| Linkin Park | 2,296,327 | Rihanna | 38,692 |
| Red Hot Chili Peppers | 2,221,660 | Adele | 38,345 |
| Lana Del Rey | 1,892,896 | David Bowie | 37,340 |
| Nirvana | 1,878,647 | Foo Fighters | 35,259 |
| System of a Down | 1,874,102 | The Killers | 35,199 |
| Florence + the Machine | 1,729,489 | Michael Jackson | 34,817 |
| Iron Maiden | 1,713,020 | The Beatles | 34,778 |
| Depeche Mode | 1,710,159 | Eminem | 34,427 |
| David Bowie | 1,685,010 | Florence + the Machine | 34,208 |
| Lady Gaga | 1,655,023 | Gorillaz | 34,157 |
| Rammstein | 1,647,437 | Pink Floyd | 33,834 |
| Queen | 1,614,548 | Linkin Park | 33,672 |
| Led Zeppelin | 1,602,110 | Katy Perry | 33,526 |
| The Black Keys | 1,517,523 | David Guetta | 33,122 |
| The xx | 1,499,181 | Maroon 5 | 32,969 |
| Nine Inch Nails | 1,489,223 | Lady Gaga | 32,728 |
| The Rolling Stones | 1,483,385 | The Cure | 32,696 |
| Eminem | 1,445,767 | Arctic Monkeys | 32,533 |
| Foo Fighters | 1,444,212 | The Doors | 32,501 |
| Arcade Fire | 1,439,458 | Johnny Cash | 32,357 |
| The Cure | 1,435,283 | U2 | 32,009 |
| Placebo | 1,427,461 | Gotye | 31,960 |

https://doi.org/10.1371/journal.pone.0217389.t001

**Table 2. Adopted music mainstreaminess measures on the user level.** Terms are explained in the text.

**Distribution-based measures**

$$M_{D,APC}^{global}(u) = 1 - avg\left(1 - \exp\left(-\Sigma_{a \in A} APC_a(u) \cdot \log \frac{APC_a(u)}{APC_a}\right), 1 - \exp\left(-\Sigma_{a \in A} APC_a \cdot \log \frac{APC_a}{APC_a(u)}\right)\right)$$

$$M_{D,APC}^{country}(u,c) = 1 - avg\left(1 - \exp\left(-\Sigma_{a \in A} APC_a(u) \cdot \log \frac{APC_a(u)}{APC_a(c)}\right), 1 - \exp\left(-\Sigma_{a \in A} APC_a(c) \cdot \log \frac{APC_a(c)}{APC_a(u)}\right)\right)$$

$$M_{D,ALC}^{global}(u) = 1 - avg\left(1 - \exp\left(-\Sigma_{a \in A} ALC_a(u) \cdot \log \frac{ALC_a(u)}{ALC_a}\right), 1 - \exp\left(-\Sigma_{a \in A} ALC_a \cdot \log \frac{ALC_a}{ALC_a(u)}\right)\right)$$

$$M_{D,ALC}^{country}(u,c) = 1 - avg\left(1 - \exp\left(-\Sigma_{a \in A} ALC_a(u) \cdot \log \frac{ALC_a(u)}{ALC_a(c)}\right), 1 - \exp\left(-\Sigma_{a \in A} ALC_a(c) \cdot \log \frac{ALC_a(c)}{ALC_a(u)}\right)\right)$$

**Rank-based measures**

$$M_{R,APC}^{global}(u) = \tau(ranks(APC), ranks(APC(u)))$$

$$M_{R,APC}^{country}(u,c) = \tau(ranks(APC(c)), ranks(APC(u)))$$

https://doi.org/10.1371/journal.pone.0217389.t002

indicates identical distributions. The formulations using the rank-based approach are based on correlations; their values are therefore in the range [−1, 1]. A value of 1 signifies that a user's artist popularity ranking is exactly the same as the one of his or her country profile (or the global profile); a value of -1 indicates exactly reverse orderings of the the user's and country's (or global) artist popularity ranking.

More precisely, in Table 2, *APC* denotes a vector containing the global artist playcounts of all artists in the dataset, keeping a fixed order (i.e., the first element in vector *APC* is the total number of listening events to the artist who is most frequently listened to globally, and so on). This vector has a size of 585,095 dimensions, one for each of the 585,095 artists in the dataset. Therefore, $APC_a$ refers to the global playcount of artist $a$. $APC_a(c)$ and $APC_a(u)$ denote, respectively, the playcount of artist $a$ on a country level (for country $c$) and on a user level (for user $u$). (Please note that, in the mainstreaminess measures on country level, country $c$ is always the country of user $u$. This may be extended in future work, to investigate, for instance, to which degree a user's music preferences overlap with the mainstream in any other country.) Likewise, $ALC_a$, $ALC_a(c)$, $ALC_a(u)$ are defined analogously, but using the listener count (instead of the playcount) of artist $a$ as popularity measure. $\tau$ denotes Kendall's rank-order correlation coefficient and *ranks* $(\cdot)$ represents the ranks of the elements in the real-valued APC or ALC vector given in $(\cdot)$, i.e., applying *ranks* $(X)$ as index on vector $X$ results in a vector that contains all elements of $X$ sorted in increasing order. Please note that Kendall's $\tau$ is undefined when using the ALC measure; therefore, this combination is not considered in our experiments. The underlying reason is that $ALC(u)$ contains only 1s for the artists listened to by $u$, irrespective of their listening frequency. Therefore, converting $ALC(u)$ values to ranks yields a single rank, which renders computing Kendall's $\tau$ impossible.

Note that higher values of mainstreaminess always indicate closer to the mainstream. Therefore, the distribution-based measures using Kullback-Leibler (KL) divergence, i.e., $M_{D,\cdot}^{country}(u, c)$ and $M_{D,\cdot}^{global}(u)$, require additional considerations. KL divergence [86] is defined on two probability distributions $P$ and $Q$, in the discrete case, as $KL(P||Q) = -\sum_x \left( P(x) \cdot log\left(\frac{P(x)}{Q(x)}\right) \right)$. We make the following adaptations: First, to normalize KL to the range [0, 1], we take $1 - \exp(-KL(P||Q))$. Second, since KL is a divergence instead of a distance metric, it is not commutative. To address this, we follow the common approach and compute the arithmetic mean (*avg* in the formulas) of $KL(P||Q)$ and $KL(Q||P)$, cf. [92, 93].

Finally, we take the inverse of the result in order to ensure that higher values indicate closer to the mainstream.

To give a few less formal examples, measure $M_{R,APC}^{country}(u, c)$ measures how well user $u$'s ranking of artist preferences, quantified in terms of artist playcounts (APC), corresponds to that of all users in country $c$. Similarly, $M_{R,ALC}^{global}(u)$ measures how well $u$'s ranking of artist preferences, quantified in terms of artist listener counts (ALC), matches with the global ranking.

### Approaches to analyze country-specific differences in music mainstream

We investigate differences in countries' music mainstream in three ways. First, we report the actual mainstreaminess values according to our measures, for all countries in the dataset, and we discuss them in the context of country-specific aspects, such as culture or market structures. Second, we uncover and discuss differences in the distribution of country-specific APC and ALC values over the global top artists. Third, we devise two outlier detection methods, implement them on the country-specific APC and ALC distributions, and discuss gained insights. Please note that we abbreviate country names according to the ISO 3166-1 alpha-2 standard (https://www.iso.org/iso-3166-country-codes.html) throughout the paper.

**Differences in mainstreaminess levels between countries.** Based on our quantitative measures of mainstreaminess introduced in Section *Approaches to measure music mainstreaminess*, (i) we compute respective values for all countries in the dataset and report descriptive statistics for each country and mainstreaminess measure. (ii) We then compare the countries and discuss the most insightful results and their meaning in the context of the country's cultural and economic background. (iii) We further report on the results of Kruskal-Wallis tests to assess whether mainstreaminess differences between countries are significant.

**Country differences in music preferences related to globally popular artists.** Based on the insights gained from the general country-specific analysis of mainstreaminess level, we deepen our investigation by illustrating and analyzing country-specific peculiarities of music preferences and of artists that form the country-specific mainstream. To this end, we create *popularity plots* that depict the APC or ALC values of the country under investigation. To uncover the extent to which country-related trends differ from the global mainstream, in the respective popularity plots, we sort artists according to their global APC or ALC values in descending order from left to right. Among the x-axis, we therefore show the artist identifiers and among the y-axis the popularity measures (APC or ALC). To illustrate such plots, Fig 1 shows global popularity curves (APC on the left side, ALC on the right side). In both plots, we see the typical, exponentially decreasing popularity values, which are indicative of a long tail characteristic [10, 54].

In order to illustrate and compare the country-specific popularity value for an artist *a* to its global popularity value, we need to rescale the global value to country *c*'s numerical range of popularity values. Eq 1 illustrates the scaling of an artist *a*'s country-specific APC value; ALC values are scaled analogously. $APC_a^{scaled(c)}$ represents the global APC value of artist *a* scaled to country *c*, *A* represents the set of all artists in the dataset, $APC_b(c)$ refers to the APC of artist *b* in country *c*, and $APC_a$ is the global APC value of artist *a*.

$$APC_a^{scaled(c)} = APC_a \cdot \frac{\Sigma_{b \in A} APC_b(c)}{\Sigma_{b \in A} APC_b} \qquad (1)$$

In the popularity plots, we display these scaled global popularity values as black curve (cf. Fig 2).

In addition, we use affinity propagation [94], which is a state-of-the-art unsupervised learning technique, to identify different categories of popularity curves. To this end, we treat the country-specific APC or ALC values of the 585,095 artists in the dataset as a feature vector and cluster them using the widely-used scikit-learn implementation of affinity propagation (http://scikit-learn.org/stable/modules/generated/sklearn.cluster.AffinityPropagation.html).

**Country-specific outlier detection and analysis.** We investigate two kinds of country-specific outliers. First, artists whose APC or ALC values in a given country substantially differ in comparison to the country-specific values of neighboring artists in terms of the global ranking. Second, artists whose APC or ALC in a given country substantially differ in comparison to the respective global value (scaled to the country's overall APC or ALC value).

To identify the former type of outliers, i.e., country-specific outliers that deviate from their neighbors in terms of APC or ALC, we use a sliding window of 5 artists, which we shift over the country-specific APC and ALC values of the top artists that are sorted in decreasing order of global popularity. Computing the mean APC and ALC value within each window and relating it to the corresponding value of the first artist *a* in the window then allows to compute the relative difference of *a*'s popularity in comparison to *a*'s neighboring artists on the artist popularity curve. If this difference exceeds a certain threshold, we consider *a* an outlier. For our experiments, we empirically set the threshold to 100%, meaning that an artist's difference to its

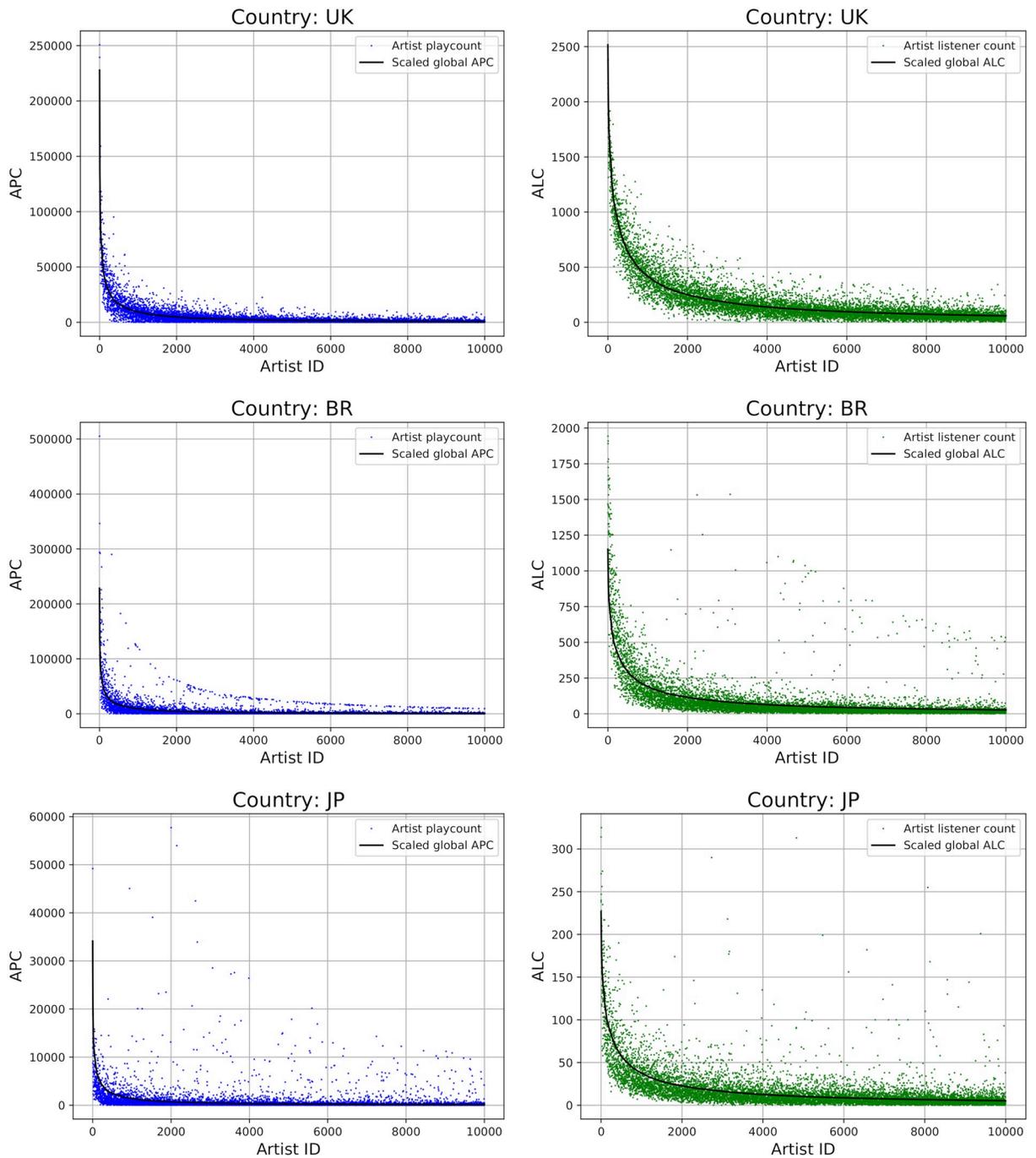

**Fig 2. Artist playcounts (APC) and artist listener counts (ALC) for the global top 10,000 artists for selected countries (UK, Brazil, and Japan).** Artist IDs (x-axis) are sorted by global popularity values.

https://doi.org/10.1371/journal.pone.0217389.g002

neighbors must be at least twice as large as the mean value in its window to qualify as a positive outlier, or at most 50% of the value of the mean value in its window to qualify as a negative outlier.

Our second approach to outlier detection quantifies the difference of an artist $a$'s APC or ALC value in the country under investigation to the respective global value. To make country-specific APC or ALC values comparable to the global values, the latter are scaled to the APC or ALC values in the respective country, as described in Section *Country differences in music preferences related to globally popular artists*. After scaling, comparing $APC_a^{scaled(c)}$ to $APC_a(c)$ allows to identify country-specific outliers that do not correspond to the global trend, meaning that the identified artists are substantially more or less popular in country *c*. In the following, we refer to the first approach to outlier detection as *sliding window approach* and to the second one as *global difference approach*.

In addition, we perform a qualitative analysis of the identified country-specific outliers, adopting both a deductive and an inductive approach. For the deductive approach, we rely on the procedure for deductive category application as used in qualitative content analysis according to Mayring [95]. We use the variables *music genre* and *gender* of the artist. For music genre, we rely on a dictionary of 20 general genres used by Allmusic (https://www.allmusic.com). For gender, we use the categories "male" and "female" for solo artists, whereas we do not consider gender for bands. Using this categorization scheme, we assign each artist identified in the quantitative outlier analysis to the corresponding category for each variable. Furthermore, we adopt an inductive approach where we rely on the inductive category development from raw data as used in qualitative content analysis according to Mayring [95]. This procedure is similar to 'open coding" within the context of Grounded Theory (cf. [96, 97]); in Mayring's approach to qualitative content analysis, the procedure for inductive category development approach is—according to Mayring [95]—more systematic compared to open coding, whose procedures are based on more rules of thumb. In the inductive category development, the authors (in their role as inductive coders) rely on their music expertise, which is, for instance, reflected in the deep engagement with fine-grained genres and music styles (as is, for example, demonstrated in a country-specific analysis of 1,998 genre and style terms retrieved from Freebase (http://www.freebase.com) [98]). Inter-coder reliability was high; and in the few instances, where some disagreement about category assignments emerged, the authors discussed the assignment in question until complete consensus could be established. Still, many categories that emerged by taking this inductive approach relate to music styles, indicating that the coding categories that emerged reflect the authors' specific expertise in music styles. Introducing additional coders with different background and expertise may potentially result in additional themes emerging as additional categories; this would, however, not affect the reliability of the category development by the existing coders. Yet, not drawing from additional (external) coders limits the breadth of categories available for analysis to what the authors have developed.

### Setup of rating prediction experiments

To compare the performance of MRS for various user groups (defined by *mainstreaminess* and *scope*, i.e., global or country), we adopt probabilistic matrix factorization for implicit feedback according to [99] for our rating prediction experiments. The reason is that music listening events can only be regarded as implicit rather than explicit feedback on user preferences. Thus, this approach is more appropriate than techniques for explicit feedback (e.g., ratings).

The model is trained through negative sampling, i.e., for each pair of user and item for which an interaction is known, another unknown pair is randomly selected to represent a negative example for training. It is noteworthy that the rating prediction experiments, we use only countries with at least 1,000 users, because only with an adequate number of users, rating prediction experiments can provide meaningful results. Thus, our experiments cover the following 13 countries: Brazil (BR), Canada (CA), Germany (DE), Spain (ES), Finland (FI), France (FR), The Netherlands (NL), Poland (PL), Russia (RU), Sweden (SE), Ukraine (UA), the United Kingdom (UK), and the United States (US). The UAM is the normalized artist playcount matrix (cf. Section *Dataset*). In 3-fold cross-validation experiments, we use root mean square error (RMSE) as performance measure, which is one of the most commonly reported error measures in recommender systems research. In each fold, we randomly split the subset of all listening events (defined by the parameters of the mainstreaminess measures: global vs. country, distribution- vs. rank-based, APC vs. ALC) into train and test set, i.e., we randomly select 80% of the listening events as training data and the remaining 20% for testing. RMSE is then averaged over the 3 runs.

RMSE is an extension to mean average error (MAE), which is defined in Eq 2, where $r_{u,i}$ and $\hat{r}_{u,i}$ respectively denote the actual and the predicted ratings of item *i* for user *u*. MAE therefore sums over the absolute prediction errors for all ratings in a test set *T* of user-item pairs. RMSE, in contrast, uses a squared error term to penalize larger discrepancies between predicted and true ratings more than smaller ones. It is defined as in Eq 3. Note that in our computation of RMSE, ratings correspond to scaled APC or ALC values.

$$MAE = \frac{1}{|T|} \sum_{r_{u,i} \in T} |r_{u,i} - \hat{r}_{u,i}| \qquad (2)$$

$$RMSE = \sqrt{\frac{1}{|T|} \sum_{r_{u,i} \in T} (r_{u,i} - \hat{r}_{u,i})^2} \qquad (3)$$

To investigate the influence of the mainstreaminess definitions (distribution- vs. rank-based and APC vs. ALC) and scopes (global vs. country-specific), cf. Table 2, as well as the mainstreaminess levels of users on recommendation performance, we then create, for each combination of mainstreaminess measure and country, respective subsets of users. More specifically, we split the users in each country into three (almost) equally sized user sets according to their mainstreaminess value: *low* corresponds to users in the lower 3-quantile (tertile) according to the respective mainstreaminess definition, *mid* and *high*, respectively, to the mid and upper tertile. We then conduct the rating prediction experiments on all users irrespective of country and mainstreaminess level (user set) as a baseline, and on the users in each user set (*low*, *mid*, and *high*) in each country, using each mainstreaminess definition. This enables a comparison of a pure mainstreaminess filtering approach (global) versus a combination of mainstreaminess filtering and country filtering (country-specific).

In other words, our overall approach roughly equals a hybrid recommendation approach in which a demographic filtering strategy is implemented before collaborative filtering is performed. The demographics in our case are defined through the different mainstreaminess groups and scopes, i.e., users are filtered according to the group and scope they are assigned to (*low*, *mid*, *high*; global or country-level). For each of the resulting user sets, we subsequently perform collaborative filtering on the group's users only.

### Validation of the results of the rating prediction experiments

We use intraclass correlation (ICC) [100] to test for consistency of the RMSE results between the 3-fold cross-validation experiments. A high degree of reliability was found between the RMSE measurements in the three runs. The average ICC was 0.999 using a 95% confidence interval.

As our data has unbalanced sample sizes across countries, we compare our experiment results gained from the full user sample with results achieved on random sub-samples of users, for which we take five times a random sample of 500 users from each country (note, in the full sample, each country has at least 1,000 users) so that we have equally-sized samples from each country. Again, we run 3-fold cross-validation experiments. We use the effect size *Cohen's d* [101] as indicator for the standardized difference between the means of the RMSE values generated when using the full sample compared to those generated when using the random subsamples. For each of the 13 countries, Cohen's *d* shows a 0 or near zero effect. Furthermore, considering Cohen's *d* for each mainstreaminess approach and user set separately also results in a 0 or near zero effect for each of the mainstreaminess approach and user set combinations. Hence, the results hold also for equal sized random samples from each country.

## Results and discussion

We present and discuss the results of our research in three subsections: (i) We provide an overview of descriptive statistics concerning the different mainstreaminess definitions that were presented in Section *Differences in mainstreaminess levels between countries* (Section *Overview of results on the mainstreaminess definitions*). (ii) We present and discuss the results of our investigations on country-specific differences of users' listening behavior concerning music mainstreaminess (Section *Country-specific music mainstreaminess*). (iii) Motivated by these results, we subsequently show how tailoring recommendations to country-specific characteristics of mainstreaminess may yield improved recommendation results (Section *Exploiting mainstream and country information for music recommendation*).

### Overview of results on the mainstreaminess definitions

Table 3 presents an overview of the descriptive statistics (mean, standard deviation, median, skewness, kurtosis, minimum, maximum, as well as first and third quartile) for all mainstream definitions, as presented in Section *Differences in mainstreaminess levels between countries*, applied to the entire dataset (i.e., users of the LFM-1b dataset with country information for countries with at least 100 users). All mainstreaminess definitions are non-normally distributed (Kolmogorov-Smirnov test, p≤0.001). The means of the APC-based definitions are about the same level for the distribution-based approach (0.150 and 0.181), and for the rank-based approach (0.171 and 0.221); with the means of the rank-based approach being slightly higher than does for the distribution-based approach. In comparison, the distribution-based definitions using ALC show higher means (0.366 and 0.446) than the APC-based definitions. The standard deviations (sd) are at similar levels for each definition (∼0.1). For the ALC-based definitions the sd is slightly higher than for the APC-based ones. While the kurtosis for the APC-based formulation in combination with the distribution-based approach is near the value of 3 that would be expected for a normal distribution, the kurtosis values for the other definitions are substantially lower, in particular for the formulation using ALC; indicating light tails, or lack of outliers.

**Table 3. Descriptive statistics for all mainstreaminess definitions for all users in the dataset with country information (47 countries with at least 100 users).**

|  | mean | sd | min | Q1 | med. | Q3 | max | skew. | kurt. |
|---|---|---|---|---|---|---|---|---|---|
| $M_{D,APC}^{global}$ | 0.150 | 0.081 | 0.004 | 0.094 | 0.139 | 0.196 | 0.886 | 1.223 | 3.111 |
| $M_{D,APC}^{country}$ | 0.180 | 0.086 | 0.006 | 0.117 | 0.167 | 0.228 | 0.997 | 1.100 | 2.800 |
| $M_{D,ALC}^{global}$ | 0.366 | 0.113 | 0.052 | 0.286 | 0.361 | 0.439 | 0.948 | 0.372 | 0.242 |
| $M_{D,ALC}^{country}$ | 0.446 | 0.118 | 0.077 | 0.363 | 0.444 | 0.525 | 0.973 | 0.155 | −0.052 |
| $M_{R,APC}^{global}$ | 0.171 | 0.099 | −0.624 | 0.115 | 0.177 | 0.234 | 0.738 | −0.538 | 1.963 |
| $M_{R,APC}^{country}$ | 0.221 | 0.110 | −0.745 | 0.158 | 0.224 | 0.288 | 1.000 | −0.164 | 2.160 |

https://doi.org/10.1371/journal.pone.0217389.t003

For the distribution-based definitions, we can clearly see that the resulting distributions of mainstreaminess values are right-skewed, indicating a generic tendency towards the inclination to mainstream; yet, the formulations with APC deliver a more pronounced picture concerning the users' indication to the mainstream than the formulations with ALC do (>1.1 vs. <0.4). For the rank-based definitions (using APC), the distributions are left-skewed.

As we will show in our recommendation experiments (Section *Exploiting mainstream and country information for music recommendation*), all these differences have an impact on recommendation performance. The level of performance depends on the user set (*low*, *mid*, and *high* mainstreaminess), though. And the direction of performance differences between the user sets depends on the mainstreaminess definitions.

### Country-specific music mainstreaminess

**Differences in mainstreaminess level.** Comparing countries, our results clearly indicate that there are country-specific differences in the level of inclination to listen to the music mainstream. In Table 4, we show descriptive statistics (mean and standard deviation) for all 47 countries in the dataset and all mainstreaminess definitions, as presented in Section *Differences in mainstreaminess levels between countries*. For every mainstreaminess measure, the countries' distributions of mainstreaminess values differ significantly (different distributions: Kruskall-Wallis test, $p \leq 0.001$; different means: Kolmogorov-Smirnov test, $p \leq 0.001$).

Depending on the mainstreaminess definition, the rank of countries with regard to the global mainstream varies. According to the distribution-based approach using APC, the country most inclined to the global mainstream is India (IN); using ALC it is Ireland (IE); according to the rank-based approach using APC it is Portugal (PT). The country most inclined to its country-specific mainstream is Estonia (EE) for the distribution-based as well as the rank-based approach, both using APC. According to the ALC-based distribution-based approach, Israel (IL) could be identified as the country most inclined to its country-specific mainstream.

Estonia (EE) and Israel (IL) appear 4 times in the top 5 country lists according to all mainstreaminess definitions. Estonia (EE) ranks first using the country-specific approaches based on APC for both, the distribution-based and the rank-based approach; with these approaches Israel (IL) ranks second. For the country-specific distribution-based approach using ALC, in contrast, Israel (IL) ranks first, followed by Estonia (EE). This indicates that users of both countries, Estonia (EE) and Israel (IL), are highly inclined to their country-specific mainstreams. Furthermore, Estonia (EE) ranks second using the rank-based approach (and APC) on a global scale; with this measure Portugal (PT) achieves a higher mainstreaminess value. Overall, users of Portugal (PT) are highly inclined to the global mainstream, as Portugal (PT) appears among the 5 countries with the highest mainstreaminess values for all 3 approaches relating to the global mainstream. In contrast, China (CN) achieves high mainstreaminess values with all 3 country-specific approaches; at the same time, China (CN) also appears in the bottom 5 country lists for all 3 mainstreaminess measures relating to the global mainstream; thus, indicating a strong inclination to the country-specific mainstream but not much for the global mainstream.

Russia (RU) is the country whose users have the lowest mainstreaminess values for both country-specific distribution-based approaches, and also having the fourth lowest values for the country-specific rank-based approach. In addition, Russia (RU) has also low mainstreaminess values when using ALC in the distribution-based approach on a global scale. Belarus (BY) has low values for both distribution-based approaches using APC and the global-scale version of the distribution-based approach using ALC. Users from the Ukraine (UA) have low mainstreaminess values when using the two country-specific distribution-based approaches; in

Table 4. **Descriptive statistics for all mainstreaminess definitions for countries with at least 100 users.** Country names are abbreviated according to ISO 3166-1 alpha-2.

| Country | No. users | $M_{D,APC}^{global}$ | | $M_{D,APC}^{country}$ | | $M_{D,ALC}^{global}$ | | $M_{D,ALC}^{country}$ | | $M_{R,APC}^{global}$ | | $M_{R,APC}^{country}$ | |
|---|---|---|---|---|---|---|---|---|---|---|---|---|---|
| | | mean | sd | mean | sd | mean | sd | mean | sd | mean | sd | mean | sd |
| AR | 282 | 0.164 | 0.090 | 0.201 | 0.087 | 0.366 | 0.118 | 0.495 | 0.105 | 0.178 | 0.117 | 0.305 | 0.124 |
| AT | 276 | 0.148 | 0.076 | 0.209 | 0.091 | 0.370 | 0.105 | 0.496 | 0.089 | 0.175 | 0.088 | 0.298 | 0.113 |
| AU | 976 | 0.157 | 0.075 | 0.187 | 0.081 | 0.384 | 0.110 | 0.455 | 0.118 | 0.175 | 0.094 | 0.235 | 0.100 |
| BE | 513 | 0.156 | 0.072 | 0.198 | 0.077 | 0.366 | 0.104 | 0.471 | 0.101 | 0.179 | 0.091 | 0.249 | 0.099 |
| BG | 236 | 0.140 | 0.074 | 0.178 | 0.078 | 0.336 | 0.101 | 0.497 | 0.076 | 0.181 | 0.102 | 0.315 | 0.107 |
| BR | 3877 | 0.169 | 0.090 | 0.182 | 0.097 | 0.383 | 0.117 | 0.435 | 0.131 | 0.184 | 0.106 | 0.230 | 0.110 |
| BY | 558 | 0.127 | 0.069 | 0.172 | 0.079 | 0.297 | 0.098 | 0.439 | 0.088 | 0.172 | 0.088 | 0.271 | 0.099 |
| CA | 1077 | 0.165 | 0.085 | 0.186 | 0.089 | 0.380 | 0.116 | 0.445 | 0.113 | 0.179 | 0.104 | 0.235 | 0.111 |
| CH | 277 | 0.143 | 0.074 | 0.202 | 0.087 | 0.366 | 0.099 | 0.504 | 0.080 | 0.157 | 0.093 | 0.279 | 0.118 |
| CL | 425 | 0.165 | 0.092 | 0.193 | 0.088 | 0.360 | 0.116 | 0.473 | 0.104 | 0.190 | 0.110 | 0.299 | 0.130 |
| CN | 162 | 0.115 | 0.075 | 0.227 | 0.091 | 0.284 | 0.116 | 0.554 | 0.078 | 0.132 | 0.108 | 0.341 | 0.135 |
| CO | 159 | 0.168 | 0.078 | 0.216 | 0.087 | 0.364 | 0.110 | 0.529 | 0.095 | 0.181 | 0.099 | 0.316 | 0.129 |
| CZ | 631 | 0.146 | 0.080 | 0.186 | 0.089 | 0.348 | 0.103 | 0.465 | 0.103 | 0.174 | 0.095 | 0.270 | 0.095 |
| DE | 4577 | 0.144 | 0.076 | 0.176 | 0.085 | 0.364 | 0.101 | 0.447 | 0.114 | 0.152 | 0.096 | 0.190 | 0.097 |
| DK | 271 | 0.146 | 0.067 | 0.184 | 0.065 | 0.376 | 0.097 | 0.510 | 0.100 | 0.169 | 0.082 | 0.274 | 0.106 |
| EE | 107 | 0.158 | 0.086 | 0.234 | 0.100 | 0.351 | 0.107 | 0.556 | 0.066 | 0.199 | 0.088 | 0.398 | 0.109 |
| ES | 1241 | 0.150 | 0.079 | 0.193 | 0.084 | 0.366 | 0.104 | 0.462 | 0.118 | 0.160 | 0.096 | 0.222 | 0.099 |
| FI | 1407 | 0.136 | 0.071 | 0.184 | 0.084 | 0.340 | 0.096 | 0.458 | 0.120 | 0.182 | 0.085 | 0.248 | 0.090 |
| FR | 1054 | 0.145 | 0.076 | 0.189 | 0.077 | 0.353 | 0.109 | 0.448 | 0.109 | 0.163 | 0.096 | 0.225 | 0.099 |
| GR | 174 | 0.156 | 0.079 | 0.223 | 0.078 | 0.357 | 0.098 | 0.531 | 0.071 | 0.167 | 0.093 | 0.317 | 0.112 |
| HR | 371 | 0.158 | 0.077 | 0.190 | 0.082 | 0.357 | 0.112 | 0.488 | 0.097 | 0.195 | 0.104 | 0.306 | 0.115 |
| HU | 272 | 0.128 | 0.065 | 0.206 | 0.080 | 0.335 | 0.103 | 0.515 | 0.078 | 0.159 | 0.108 | 0.308 | 0.127 |
| ID | 484 | 0.145 | 0.086 | 0.214 | 0.095 | 0.343 | 0.111 | 0.492 | 0.114 | 0.142 | 0.101 | 0.253 | 0.121 |
| IE | 220 | 0.179 | 0.085 | 0.212 | 0.082 | 0.415 | 0.115 | 0.499 | 0.097 | 0.189 | 0.102 | 0.301 | 0.126 |
| IL | 100 | 0.171 | 0.087 | 0.232 | 0.117 | 0.364 | 0.112 | 0.569 | 0.069 | 0.187 | 0.114 | 0.392 | 0.153 |
| IN | 122 | 0.188 | 0.113 | 0.229 | 0.112 | 0.377 | 0.133 | 0.545 | 0.086 | 0.168 | 0.135 | 0.336 | 0.144 |
| IR | 135 | 0.129 | 0.081 | 0.202 | 0.087 | 0.334 | 0.116 | 0.550 | 0.089 | 0.130 | 0.130 | 0.327 | 0.149 |
| IT | 972 | 0.158 | 0.080 | 0.196 | 0.087 | 0.370 | 0.105 | 0.460 | 0.118 | 0.167 | 0.099 | 0.242 | 0.105 |
| JP | 804 | 0.105 | 0.072 | 0.190 | 0.087 | 0.254 | 0.110 | 0.480 | 0.100 | 0.165 | 0.104 | 0.259 | 0.110 |
| LT | 202 | 0.153 | 0.078 | 0.205 | 0.087 | 0.339 | 0.100 | 0.509 | 0.080 | 0.183 | 0.080 | 0.328 | 0.097 |
| LV | 165 | 0.147 | 0.080 | 0.191 | 0.071 | 0.355 | 0.117 | 0.531 | 0.082 | 0.182 | 0.081 | 0.327 | 0.107 |
| MX | 705 | 0.157 | 0.084 | 0.189 | 0.089 | 0.348 | 0.103 | 0.452 | 0.116 | 0.172 | 0.102 | 0.252 | 0.106 |
| NL | 1375 | 0.155 | 0.076 | 0.180 | 0.081 | 0.377 | 0.106 | 0.447 | 0.114 | 0.185 | 0.097 | 0.227 | 0.098 |
| NO | 749 | 0.142 | 0.067 | 0.176 | 0.077 | 0.370 | 0.092 | 0.470 | 0.113 | 0.171 | 0.085 | 0.231 | 0.090 |
| NZ | 163 | 0.168 | 0.086 | 0.216 | 0.082 | 0.392 | 0.116 | 0.539 | 0.085 | 0.185 | 0.099 | 0.326 | 0.118 |
| PL | 4402 | 0.156 | 0.084 | 0.171 | 0.089 | 0.368 | 0.113 | 0.427 | 0.121 | 0.187 | 0.099 | 0.231 | 0.101 |
| PT | 291 | 0.182 | 0.088 | 0.209 | 0.085 | 0.389 | 0.117 | 0.503 | 0.098 | 0.200 | 0.102 | 0.308 | 0.125 |
| RO | 237 | 0.143 | 0.077 | 0.189 | 0.084 | 0.343 | 0.103 | 0.516 | 0.087 | 0.190 | 0.086 | 0.312 | 0.101 |
| RS | 253 | 0.145 | 0.078 | 0.204 | 0.079 | 0.340 | 0.104 | 0.515 | 0.077 | 0.190 | 0.095 | 0.323 | 0.116 |
| RU | 5013 | 0.130 | 0.075 | 0.154 | 0.078 | 0.305 | 0.102 | 0.392 | 0.102 | 0.166 | 0.094 | 0.201 | 0.094 |
| SE | 1231 | 0.136 | 0.064 | 0.165 | 0.073 | 0.366 | 0.094 | 0.463 | 0.115 | 0.169 | 0.083 | 0.214 | 0.088 |
| SK | 192 | 0.153 | 0.076 | 0.219 | 0.096 | 0.361 | 0.104 | 0.546 | 0.077 | 0.178 | 0.091 | 0.336 | 0.113 |
| TR | 479 | 0.158 | 0.083 | 0.186 | 0.084 | 0.370 | 0.106 | 0.490 | 0.115 | 0.166 | 0.092 | 0.264 | 0.101 |
| UA | 1142 | 0.137 | 0.076 | 0.165 | 0.080 | 0.311 | 0.105 | 0.415 | 0.101 | 0.165 | 0.091 | 0.235 | 0.098 |
| UK | 4533 | 0.169 | 0.081 | 0.179 | 0.085 | 0.404 | 0.113 | 0.437 | 0.120 | 0.177 | 0.098 | 0.193 | 0.100 |

(Continued)

Table 4. (Continued)

| Country | No. users | $M_{D.APC}^{global}$ | | $M_{D.APC}^{country}$ | | $M_{D.ALC}^{global}$ | | $M_{D.ALC}^{country}$ | | $M_{R.APC}^{global}$ | | $M_{R.APC}^{country}$ | |
|---|---|---|---|---|---|---|---|---|---|---|---|---|---|
| | | mean | sd | mean | sd | mean | sd | mean | sd | mean | sd | mean | sd |
| US | 10248 | 0.162 | 0.082 | 0.177 | 0.086 | 0.397 | 0.112 | 0.438 | 0.120 | 0.162 | 0.101 | 0.177 | 0.101 |
| VE | 118 | 0.167 | 0.085 | 0.227 | 0.086 | 0.367 | 0.114 | 0.541 | 0.088 | 0.187 | 0.099 | 0.372 | 0.138 |
| Total | 53258 | 0.152 | 0.081 | 0.180 | 0.086 | 0.366 | 0.113 | 0.446 | 0.118 | 0.171 | 0.099 | 0.221 | 0.110 |

https://doi.org/10.1371/journal.pone.0217389.t004

addition, also with the distribution-based approach using ALC on a global scale they have low mainstreaminess values. Iran (IR) appears 2 times in the bottom 5 country lists, namely for both approaches using APC and relating to the global mainstream. Using the country-specific distribution-based approach using ALC, Iran (IR) is among the countries with the highest mainstreaminess values. While Russia (RU) scores lowest for both country-specific distribution-based approaches (and, in addition, it has the fourth-lowest score with the country-specific rank-based approach), whereas Japan (JP) scores lowest for both global-scale distribution-based approaches. In other words, our results suggest that users from Russia (RU) are not inclined to a "Russian mainstream" and users from Japan (JP) are the least inclined to the global mainstream. The Unites States (US) score lowest when using the rank-based approach for country-specific mainstream; the United States (US) does, though, not appear in any other of the top 5 or bottom 5 country lists.

As some countries rank high for both, the global and the country-specific mainstreaminess definition, we next inspect details concerning country-specific differences of music preferences related to globally popular artists, which we present in Section *Differences in music preferences related to globally popular artists*, and country-specific outliers, which we present in Section *Country-specific outliers*.

**Differences in music preferences related to globally popular artists.** Adopting the approach described in Section *Country differences in music preferences related to globally popular artists*, we plot popularity curves and depict in Fig 2 the results for three selected countries: the United Kingdom (UK), Brazil (BR), and Japan (JP), with respect to APC (plots on the left side) and ALC (plots on the right side). We selected these three countries after careful inspection of all plots, because they are representative for the three archetypes we empirically identified: (i) those countries where the mainstream of the country largely corresponds to the global trend (in addition to the United Kingdom (UK), also the United States (US) and the Netherlands (NL) fall into this category, among others), (ii) those countries with a distinct country-specific mainstream in addition to the global mainstream (e.g., Brazil (BR), Russia (RU), and Finland (FI)), and (iii) those countries roughly following the global mainstream, but at the same time showing various country-specific outliers over the whole global artist popularity range (e.g., Japan (JP), China (CN), and Indonesia (ID)). By definition, global mainstreaminess may be more intensively shaped by countries with more users. However, this does not necessarily affect the effectiveness of the proposed measures in MRS. Results from experiments with a sub-sample (cf. Section *Validation of the results of the rating prediction experiments*), where 500 users were randomly selected from each country (five times each), delivered similar results to using the complete user sample.

To investigate whether this manual categorization is reflected in an automatically generated clustering of countries, we next apply affinity propagation [94] as described in Section *Country differences in music preferences related to globally popular artists*. Using an Euclidean affinity definition, a damping factor of 0.5, and a maximum of 200 iterations, affinity propagation

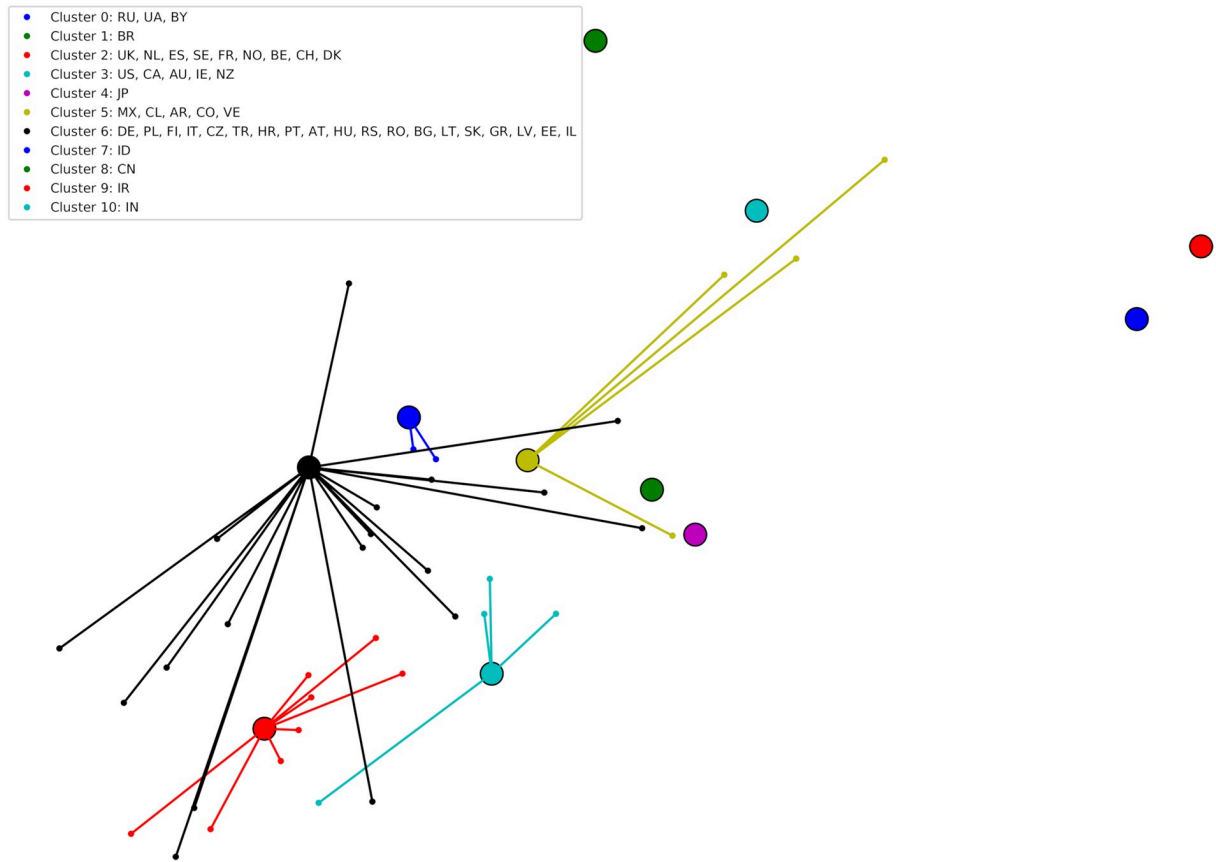

**Fig 3. Affinity propagation results on the ALC values and countries assigned to each cluster.**

https://doi.org/10.1371/journal.pone.0217389.g003

detects 11 clusters in the ALC data. Fig 3 shows the resulting clustering of countries, where items detected as prototypes, i.e., cluster centers, are visualized as larger circles, and connected to the other items belonging to the same cluster. Fig 4 depicts the corresponding cluster centers or prototypes. As we can see in Fig 4, Clusters 2, 3, and 6 show relatively few strong outliers. They therefore represent countries whose mainstream largely corresponds to the global mainstream. These clusters correspond to category (i) found in our empirical observation above. On the other hand, we observe clusters such as Cluster 0 and 1, whose countries developed a distinct country-specific mainstream in parallel to the global one. This is indicated by the observable second curve above the main distribution in Cluster 0 and less pronounced in Cluster 1. These clusters correspond to category (ii) of our empirical observation. Finally, affinity propagation also detected several countries whose artist popularity distribution is scattered across the global distribution, therefore indicating various country-specific outliers. These correspond to group (iii) of our empirical categorization. Since the structure of these outliers depend on the country each forms a cluster on its own in Fig 3.

Looking at the country composition of clusters (cf. Fig 3), we observe strong relationships in the cultural, historic, and linguistic background of countries in the same cluster. For instance, Cluster 3 contains only countries in which English is the main language. Cluster 0 comprises countries of the former Soviet Union, thus sharing a joint historic background. Cluster 5 includes all South American countries whose language is Spanish, while Portuguese-speaking Brazil (BR) forms a cluster on its own (Cluster 1). Clusters 2 and 6 contain European

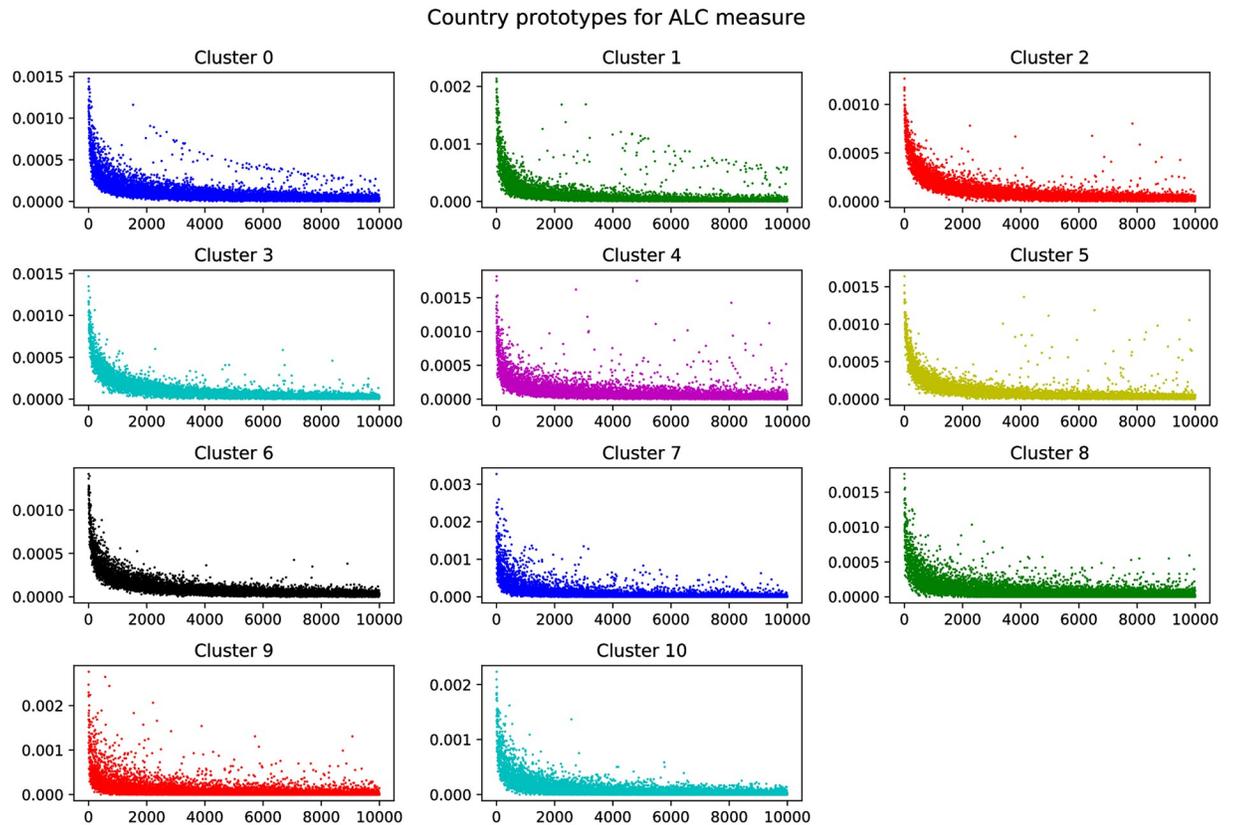

**Fig 4. Cluster prototypes resulting from affinity propagation on the ALC values.**

https://doi.org/10.1371/journal.pone.0217389.g004

countries. Interestingly, countries in Cluster 2 are foremost Western and Northern European countries, whereas, except for a few outliers, Cluster 6 represents Central, Eastern, and Southern European countries. Cluster 2 contains the strong music nations in terms of popular music creation and production: the United Kingdom (UK), the Netherlands (NL), Sweden (SE), and France (FR). In addition, Belgium (BE) and Switzerland (CH) are part of Cluster 2, likely because of (partly) overlapping languages of their citizens (France (FR) and the Netherlands (NL) that both also belong to Cluster 2). Cluster 6, on the other hand, contains countries that share long borders (e.g., Germany (DE) and Poland (PL)) or that can look back to a joint history (e.g., the former Austrian-Hungarian monarchy or the Baltic countries). Countries with a mainstream substantially far away from the global one form clusters of their own, e.g., Japan (JP) (Cluster 4), Indonesia (ID) (Cluster 7), China (CN) (Cluster 8), Iran (IR) (Cluster 9), and India (IN) (Cluster 10).

**Country-specific outliers.** Using the approach described in Section *Country-specific outlier detection and analysis*, we set out to identify two kinds of outliers: artists whose APC or ALC value differs from that of their neighbors in the popularity curve, and artists whose APC or ALC value differs from the respective global value. We consider the top 10,000 global artists to compute country-specific outliers from the mainstream. In the following, we select four countries which we investigate further in terms of outliers, i.e., the United Kingdom (UK), Finland (FI), Brazil (BR), and Japan (JP), because they represent different archetypes of country-specific mainstream evolution, cf. Section *Differences in music preferences related to globally popular artists*.

**Table 5. Top 20 positive (top) and negative (bottom) outliers for the United Kingdom (UK).** Differences to the mainstream are given according to the sliding window approach and the global difference approach, using APC to measure popularity.

| Artist | Global rank | Sliding window | Global difference |
| --- | --- | --- | --- |
| Neutral Milk Hotel | 348 | +120.06 | +174.34 |
| Biffy Clyro | 377 | +202.74 | +480.65 |
| Kate Bush | 413 | +149.78 | +318.84 |
| Manic Street Preachers | 430 | +179.11 | +439.35 |
| The Weeknd | 441 | +104.66 | +219.61 |
| Four Tet | 489 | +108.27 | +243.39 |
| Jeff Buckley | 498 | +123.75 | +192.33 |
| Lostprophets | 536 | +107.47 | +211.49 |
| Pulp | 607 | +111.82 | +327.49 |
| The Vaccines | 629 | +126.97 | +224.40 |
| The Stone Roses | 640 | +195.28 | +314.48 |
| Elbow | 647 | +223.15 | +446.81 |
| Fugazi | 649 | +152.36 | +175.70 |
| Frank Turner | 676 | +199.45 | +458.22 |
| You Me at Six | 729 | +142.90 | +359.74 |
| Angels & Airwaves | 756 | +109.05 | +134.00 |
| McFly | 798 | +159.79 | +454.35 |
| Purity Ring | 805 | +131.80 | +137.27 |
| MF DOOM | 814 | +138.20 | +195.19 |
| Delain | 815 | +114.82 | +112.60 |
| Katatonia | 91 | -56.66 | -62.60 |
| Anathema | 113 | -52.41 | -49.87 |
| God Is An Astronaut | 141 | -55.36 | -54.26 |
| Children of Bodom | 153 | -51.70 | -55.38 |
| Three Days Grace | 156 | -61.90 | -63.28 |
| Norah Jones | 162 | -50.30 | -40.75 |
| Apocalyptica | 184 | -63.56 | -64.18 |
| Eluveitie | 193 | -61.62 | -55.86 |
| Mariah Carey | 236 | -52.52 | -51.76 |
| Two Steps from Hell | 242 | -56.05 | -65.94 |
| Infected Mushroom | 243 | -50.85 | -50.29 |
| Asking Alexandria | 254 | -51.81 | -37.79 |
| Helloween | 266 | -59.47 | -61.57 |
| fun. | 279 | -50.62 | -47.36 |
| Volbeat | 310 | -55.03 | -51.94 |
| Skillet | 318 | -74.42 | -69.88 |
| Armin van Buuren | 320 | -57.73 | -50.20 |
| Archive | 335 | -63.89 | -71.95 |
| Coma | 339 | -87.81 | -89.42 |
| Sting | 352 | -57.76 | -48.13 |

https://doi.org/10.1371/journal.pone.0217389.t005

Table 5 shows the outliers for the *United Kingdom (UK)*. The first thing we notice is that the first positive outlier does not occur before rank 348 in the global ranking, which indicates that listeners from the United Kingdom do not reveal an exorbitant preference for any of the globally top artists, thus corresponding to archetype (i) identified in Section *Differences in music preferences related to globally popular artists*. This is in stark contrast to the other three countries we investigate in detail, in which the first positive outlier can be found at much lower global ranks: 1, 19, 48, respectively, for Japan (JP), Finland (FI), and Brazil (BR).

Among the negative outliers in the United Kingdom, we predominantly find metal and hard rock bands, e.g., Katatonia, Children of Bodom, Apocalyptica, Eluveitie, and many more.

Big differences between the sliding window approach and the global difference approach can be found for bands such as Biffy Clyro (203% vs. 481%) and McFly (160% vs. 454%), the former being a Scottish rock band, the latter a pop band from London. Here we can nicely observe that the global difference definition particularly highlights bands that originate from the country under investigation, while the sliding window approach uncovers local trends that do not comply with the country-specific popularity pattern.

Next, we investigate Finland (FI) and Brazil (BR), archetypes for a category of countries with a distinct country-specific mainstream in addition to the global mainstream, cf. (ii) identified in Section *Differences in music preferences related to globally popular artists*. As for outliers in *Finland (FI)*, Table 6 reveals that the positive outliers are almost exclusively composed of metal bands, with Rammstein, In Flames, and Megadeth occupying the top five places. Similar to our observation for the United Kingdom (UK), the top outlier according to the global difference approach is a Finnish band (Amorphis with 724%). Interestingly, the second and third places are taken by an Isreali and a US band, respectively, Infected Mushrom with 488% and Lamb of God with 398%.

Negative outliers among Finnish listeners include artists of a variety of music styles, with a particularly low popularity of Canadian rapper Drake: −92% and −90% for the sliding window a global difference approach, respectively.

Listeners in *Brazil (BR)*, whose outliers are shown in Table 7, seem to prefer female pop singers more than the global average. Among the top outliers we find, for instance, Britney Spears, Beyoncé, Avril Lavigne, Kylie Minogue, Christina Aguilera, and P!nk. Particularly negative outliers are foremost composed of electronic music artists, such as Boards of Canada (−83% with respect to global difference approach), Pendulum (−80%), and Bonobo (−78%).

Table 8 shows the dominant outliers for *Japan (JP)*. Here, we find the only classical music composer, Wolfgang Amadeus Mozart, occurring among the top 20 outliers of any of the investigated countries. Indeed, Mozart is played almost twice as often in Japan (JP) than globally. In contrast, musical styles like indie, alternative, and progressive rock are seemingly rather unpopular in Japan (JP), evidenced by the bands The Black Keys (−90% according to global difference approach), Rise Against (−86%), and Florence + the Machine (−79%). Also American pop singer and songwriter Lana Del Rey (−85%) and Swedish metal band Katatonia (−84%) range among the most negative outliers in Japan.

## Exploiting mainstream and country information for music recommendation

Based on the results presented in Section *Country-specific music mainstreaminess*, we now investigate whether users are better served with a state-of-the-art collaborative filtering MRS that tailors its recommendations based on nearest neighbors drawn from the entire user set or, in comparison, by a system that considers nearest neighbors only among users with similar global or country-specific mainstreaminess. We therefore analyze the influence of filtering the nearest neighbors of the target user with respect to the mainstreaminess group (*low*, *mid*, or *high*) they belong to (cf. Section *Setup of rating prediction experiments*).

Table 9 summarizes the root mean square error (RMSE) for the various global and country-specific mainstreaminess definitions, averaged over all considered countries.

**Table 6. Top 20 positive (top) and negative (bottom) outliers for Finland (FI).** Differences to the mainstream are given according to the sliding window approach and the global difference approach, using APC to measure popularity.

| Artist | Global rank | Sliding window | Global difference |
| --- | --- | --- | --- |
| Rammstein | 19 | +113.94 | +175.41 |
| In Flames | 38 | +167.97 | +319.86 |
| Megadeth | 55 | +112.92 | +219.59 |
| Katatonia | 91 | +160.51 | +270.90 |
| Pendulum | 102 | +155.95 | +257.48 |
| Amon Amarth | 115 | +100.49 | +239.22 |
| Bullet for My Valentine | 148 | +161.63 | +195.53 |
| Children of Bodom | 153 | +166.83 | +335.39 |
| Sonata Arctica | 168 | +192.31 | +345.43 |
| Sum 41 | 176 | +107.90 | +109.71 |
| HIM | 197 | +146.16 | +280.88 |
| Rush | 199 | +103.88 | +129.65 |
| Lamb of God | 208 | +174.11 | +397.66 |
| Sabaton | 226 | +150.89 | +287.51 |
| Infected Mushroom | 243 | +205.13 | +487.54 |
| Amorphis | 265 | +211.43 | +724.49 |
| Ensiferum | 287 | +123.93 | +340.63 |
| Korpiklaani | 302 | +143.69 | +241.22 |
| Gojira | 313 | +134.04 | +312.04 |
| Five Finger Death Punch | 314 | +184.74 | +238.13 |
| Adele | 34 | -63.17 | -45.80 |
| The Killers | 35 | -61.17 | -38.86 |
| Beyoncé | 75 | -56.83 | -66.15 |
| John Mayer | 94 | -64.98 | -74.75 |
| Maroon 5 | 96 | -60.41 | -59.70 |
| Vampire Weekend | 105 | -51.85 | -75.97 |
| Jack Johnson | 109 | -70.74 | -75.79 |
| Portishead | 112 | -50.74 | -43.47 |
| Beck | 123 | -54.08 | -64.67 |
| Incubus | 131 | -62.86 | -66.20 |
| Elliott Smith | 147 | -65.73 | -75.74 |
| Kasabian | 150 | -65.18 | -64.72 |
| Kylie Minogue | 151 | -55.16 | -48.36 |
| Jay-Z | 152 | -67.06 | -59.39 |
| Animal Collective | 161 | -63.63 | -77.60 |
| The Shins | 163 | -51.36 | -68.00 |
| Drake | 166 | -91.62 | -90.48 |
| Beach House | 170 | -57.37 | -71.13 |
| Fleet Foxes | 179 | -53.47 | -70.18 |
| Fleetwood Mac | 192 | -54.64 | -65.87 |

https://doi.org/10.1371/journal.pone.0217389.t006

On average, compared to a standard collaborative filtering approach operating on the entire user set ($RMSE$ = 218.137), considering a user's mainstreaminess level delivers substantially improved results for each of the 6 proposed mainstreaminess definitions (with RMSE ranging around ∼25.3 to ∼26.2, and an average over all mainstreaminess definitions of $RMSE$ = 25.268). At the same time, on average, all 6 proposed mainstreaminess definitions

**Table 7. Top 20 positive (top) and negative (bottom) outliers for Brazil (BR).** Differences to the mainstream are given according to the sliding window approach and the global difference approach, using APC to measure popularity.

| Artist | Global rank | Sliding window | Global difference |
|---|---|---|---|
| The Strokes | 48 | +145.66 | +332.62 |
| Britney Spears | 57 | +180.13 | +425.25 |
| Dream Theater | 62 | +101.63 | +211.36 |
| Beyoncé | 75 | +109.39 | +346.95 |
| Avril Lavigne | 100 | +141.77 | +334.23 |
| Ramones | 132 | +137.23 | +316.71 |
| Pantera | 140 | +118.37 | +183.18 |
| Kylie Minogue | 151 | +145.38 | +269.24 |
| Panic! at the Disco | 186 | +108.14 | +329.16 |
| Rush | 199 | +106.47 | +198.14 |
| Christina Aguilera | 213 | +177.45 | +441.10 |
| Cat Power | 214 | +119.75 | +177.65 |
| P!nk | 220 | +149.86 | +271.00 |
| Mariah Carey | 236 | +106.21 | +197.91 |
| Bruno Mars | 237 | +146.76 | +169.88 |
| Enya | 238 | +116.29 | +82.30 |
| Kiss | 278 | +113.27 | +274.91 |
| Garbage | 308 | +209.27 | +251.46 |
| Lily Allen | 321 | +110.29 | +224.15 |
| Miley Cyrus | 333 | +141.67 | +377.06 |
| Depeche Mode | 16 | -52.49 | -33.21 |
| Nine Inch Nails | 24 | -55.34 | -48.77 |
| In Flames | 38 | -63.70 | -65.69 |
| Kanye West | 40 | -62.40 | -47.04 |
| The National | 42 | -72.59 | -58.67 |
| Massive Attack | 52 | -67.76 | -69.44 |
| Sigur Rós | 54 | -66.32 | -42.15 |
| Bon Iver | 56 | -57.74 | -36.49 |
| Boards of Canada | 63 | -80.50 | -82.65 |
| Tool | 66 | -69.84 | -53.98 |
| Bonobo | 68 | -88.78 | -77.83 |
| The Prodigy | 76 | -79.18 | -76.08 |
| M83 | 77 | -78.34 | -68.57 |
| Moby | 92 | -75.05 | -63.24 |
| Pendulum | 102 | -75.29 | -80.26 |
| Modest Mouse | 103 | -50.92 | -45.06 |
| Vampire Weekend | 105 | -58.12 | -32.76 |
| Röyksopp | 126 | -78.77 | -76.41 |
| Sufjan Stevens | 127 | -75.11 | -65.73 |
| Bruce Springsteen | 133 | -50.56 | -59.79 |
| deadmau5 | 143 | -60.92 | -68.72 |
| Bloc Party | 146 | -52.86 | -50.97 |

https://doi.org/10.1371/journal.pone.0217389.t007

deliver comparable results. The results suggest that the country-specific definitions deliver an average RMSE (*RMSE* = 25.642) comparable to that achieved with the global mainstream definitions (25.721). The approaches using APC deliver slightly lower average RMSE (25.469) than those using ALC (*RMSE* = 26.104).

**Table 8. Top 20 positive (top) and negative (bottom) outliers for Japan (JP).** Differences to the mainstream are given according to the sliding window approach and the global difference approach, using APC to measure popularity.

| Artist | Global rank | Sliding window | Global difference |
|---|---|---|---|
| The Beatles | 1 | +109.55 | +155.85 |
| Green Day | 33 | +103.97 | +136.74 |
| Sigur Rós | 54 | +101.18 | +161.63 |
| Boards of Canada | 63 | +113.03 | +139.90 |
| Oasis | 71 | +163.20 | +165.89 |
| Björk | 89 | +130.97 | +177.07 |
| Avril Lavigne | 100 | +124.48 | +201.71 |
| Mogwai | 137 | +131.85 | +162.72 |
| Norah Jones | 162 | +129.70 | +191.15 |
| Aphex Twin | 169 | +132.03 | +221.91 |
| Miles Davis | 191 | +153.25 | +303.63 |
| The Chemical Brothers | 223 | +278.73 | +242.12 |
| Metric | 224 | +102.69 | +46.26 |
| Enter Shikari | 232 | +196.14 | +196.67 |
| Burial | 240 | +133.29 | +125.50 |
| Wolfgang Amadeus Mozart | 259 | +135.34 | +191.59 |
| Flying Lotus | 275 | +225.78 | +261.58 |
| My Bloody Valentine | 283 | +229.97 | +241.40 |
| Morrissey | 306 | +194.99 | +231.43 |
| The Flaming Lips | 330 | +169.75 | +169.46 |
| Lana Del Rey | 11 | -54.21 | -85.23 |
| Florence + the Machine | 14 | -62.63 | -78.94 |
| The Black Keys | 22 | -82.64 | -90.06 |
| Placebo | 30 | -62.78 | -68.40 |
| Mumford & Sons | 50 | -74.97 | -82.05 |
| blink-182 | 53 | -59.36 | -69.38 |
| Rise Against | 61 | -78.68 | -86.21 |
| Johnny Cash | 65 | -62.30 | -81.55 |
| Bonobo | 68 | -52.36 | -65.84 |
| Beyoncé | 75 | -58.68 | -72.38 |
| Katatonia | 91 | -69.55 | -84.48 |
| 30 Seconds to Mars | 98 | -67.88 | -75.94 |
| Porcupine Tree | 99 | -51.36 | -62.42 |
| Modest Mouse | 103 | -59.32 | -67.36 |
| Amon Amarth | 115 | -56.35 | -74.20 |
| Yann Tiersen | 119 | -66.53 | -78.52 |
| Hans Zimmer | 142 | -64.66 | -67.73 |
| deadmau5 | 143 | -59.38 | -60.11 |
| Jay-Z | 152 | -50.48 | -70.62 |
| Three Days Grace | 156 | -73.76 | -77.34 |

https://doi.org/10.1371/journal.pone.0217389.t008

Further analysis shows that for the definitions considering mainstreaminess on a global scale, the average RMSE is lower when using APC ($RMSE = 25.437$) than for using ALC ($RMSE = 26.048$). For the country-specific mainstreaminess definitions, in contrast, RMSE is lower when using ALC ($RMSE = 25.501$) compared to using APC ($RMSE = 26.159$). A comparison between the APC- and the ALC-based definitions shows that for both, APC and ALC,

Table 9. Root mean square error (RMSE) for the various components of our mainstreaminess definitions (distribution- vs. rank-based, APC vs. ALC, and global vs. country scope), averaged over all considered countries and user sets.

| Type of mainstreaminess approach | | | RMSE |
|---|---|---|---|
| No mainstreaminess approach (i.e., baseline) | | | 218.137 |
| Mean over distribution-based approaches | | | 25.865 |
| Mean over rank-based approaches | | | 25.310 |
| Mean over APC-based approaches | | | 25.469 |
| Mean over ALC-based approaches | | | 26.104 |
| Mean over global-scope approaches | | | 25.642 |
| Mean over country-scope approaches | | | 25.721 |
| Distribution-based approaches | APC-based approaches | $M_{D,APC}^{global}$ | 25.591 |
| | | $M_{D,APC}^{country}$ | 25.662 |
| | ALC-based approaches | $M_{D,ALC}^{global}$ | 26.048 |
| | | $M_{D,ALC}^{country}$ | 26.159 |
| Rank-based approaches | APC-based approaches | $M_{R,APC}^{global}$ | 25.279 |
| | | $M_{R,APC}^{country}$ | 25.341 |
| Mean over all mainstreaminess definitions | | | 25.682 |

https://doi.org/10.1371/journal.pone.0217389.t009

there is almost no difference between considering the country-specific mainstream compared to considering the global mainstream (global mainstream *RMSE* = 25.437 vs. country-specific mainstream *RMSE* = 25.501 for APC; and global mainstream 26.048 vs. country-specific mainstream *RMSE* = 26.159 for ALC). The ALC-based definitions, though, deliver higher average RMSE (*RMSE* = ∼25.5) for both, considering a country-specific or a globally-defined mainstreaminess definition, compared to the APC-based definitions (*RMSE* = ∼26.1).

Table 10 shows the RMSE for the global and country-specific mainstreaminess definitions and various levels of mainstreaminess (i.e., user sets), averaged over all considered countries.

Overall, compared to the various mainstreaminess approaches and user sets, the lowest RMSE is achieved for the *low* user set when using the $M_{R,APC}^{country}$ approach (*RMSE* = 20.423). Interestingly, the worst result is achieved for the *high* user set with the very same approach (*RMSE* = 28.742). Accordingly, this approach is also the one with the most discrepancies across the three user sets. A high span across the results of the three user sets is also present using the $M_{R,APC}^{global}$ approach (*RMSE* = 20.578 for *low* and 28.644 for *high*).

Irrespective of the scope (global or country-specific), results for the distribution-based approaches suggest that using APC for the mainstreaminess definition achieves the best results for the *low* user set (*RMSE* = 23.824 for global and 23.694 for country-specific), the worst for the *mid* user set (*RMSE* = 26.712 and 26.688, respectively), and the results for the *high* user set (*RMSE* = 26.239 and 26.604) are only slightly better than the ones for the *mid* user set (*RMSE* = 26.712 and *RMSE* = 26.688). In contrast, using ALC, the *high* user set is served best (*RMSE* = 25.139 for the global and 25.263 for the country-specific version); merely slightly worse RMSE values are achieved for the *low* user set (*RMSE* = 25.688 and 26.098), and considerably worse for the *mid* user set (*RMSE* = 27.315 versus 27.117).

Compared to the standard collaborative filtering approach on the entire user set, the strongest improvement is achieved for the *high* user set when using the $M_{D,ALC}^{global}$ approach (*RMSE* = 25.139). For the *mid* user set, the lowest RMSE value (26.688) is achieved with $M_{D,APC}^{country}$. The *low* user set is best served with the $M_{R,APC}^{country}$ definition (*RMSE* = 20.423).

**Table 10. Root mean square error (RMSE) for the global and country-specific mainstreaminess definitions and various levels of mainstreaminess, i.e. user sets, averaged over all considered countries.**

| Type of mainstreaminess approach | | | No. users | User set | RMSE |
|---|---|---|---|---|---|
| No mainstreaminess approach (i.e., baseline) | | | 53,259 | — | 218.137 |
| distribution-based approaches | APC-based approaches | $M_{D,APC}^{global}$ | each user set 17,753 | high | 26.239 |
| | | | | mid | 26.712 |
| | | | | low | 23.824 |
| | | $M_{D,APC}^{country}$ | | high | 26.604 |
| | | | | mid | 26.688 |
| | | | | low | 23.694 |
| | ALC-based approaches | $M_{D,ALC}^{global}$ | | high | 25.139 |
| | | | | mid | 27.315 |
| | | | | low | 25.688 |
| | | $M_{D,ALC}^{country}$ | | high | 25.263 |
| | | | | mid | 27.117 |
| | | | | low | 26.098 |
| rank-based approaches | APC-based approaches | $M_{R,APC}^{global}$ | | high | 28.644 |
| | | | | mid | 26.726 |
| | | | | low | 20.578 |
| | | $M_{R,APC}^{country}$ | | high | 28.742 |
| | | | | mid | 26.858 |
| | | | | low | 20.423 |

https://doi.org/10.1371/journal.pone.0217389.t010

Typically, it is particularly difficult to predict the preferences and listening behavior of users with a highly specialized music taste; thus, in terms of mainstreaminess, it is expected that recommendations for the *low* user set would result in a higher average RMSE than for the other user sets, while the *high* user set would render lower average RMSE values. However, considering only users with similar mainstreaminess levels (differentiating three user segments with *low*, *mid*, and *high* mainstreaminess levels, respectively)—as we do in our proposed approaches—seems to serve the *low* user segment particularly well: compared to the *high* and *mid* user sets, the *low* user set achieves the lowest average RMSE for all APC-based approaches. With the ALC-based approaches, in contrast, the *high* user set is served slightly better than the *low* segment. The results for the particularly difficult *low* user set for the APC-based mainstreaminess definitions are even below the average across all mainstreaminess definitions and user sets (*RMSE* = 25.682). for $M_{R,APC}^{global}$ and $M_{R,APC}^{country}$, respectively, the rank-based approaches seem to work particularly well for the *low* mainstreaminess users. Beyond that, these average RMSE values are by far the lowest ones achieved across all mainstreaminess definitions and user sets. Overall, these results strongly suggest that such a separation of users is beneficial for users with specialized music tastes, regardless of whether mainstreaminess is defined on the country or a global level. The only exception is $M_{D,ALC}^{country}$, where the RMSE for the *low* user segment (*RMSE* = 26.098) is slightly higher than the average RMSE across all mainstreaminess definitions (*RMSE* = 25.682).

Moreover, for the *low* user set, the RMSE is generally considerably higher for the ALC-based mainstreaminess definitions (*RMSE* = 25.688 for global and 26.098 for country-specific) compared to the APC-based definitions (distribution-based approaches: global *RMSE* = 23.824 and country-specific *RMSE* = 23.694; rank-based approaches: global 20.578 and country-specific 20.423). This can be explained as follows. ALC computed on the user level results in a binary representation of the user's listening behavior, i.e., 1 if the user listened at least once to the artist and 0 otherwise. This implies a loss of listening frequency information. As a result, information encoded in the measure of artist popularity about long tail artists (which are the artists listened to by low mainstreaminess listeners) is further reduced.

## Conclusion and further research

Popularity-based approaches are widely adopted in music recommendation systems, both in industry and research. However, as the popularity distribution of music items is typically a long-tail distribution, current approaches to music recommendations fall short in satisfying listeners that have specialized music preferences far away from the global music mainstream.

Most research on music recommender systems falls short in considering country-specific differences of popularity. Furthermore, the approaches currently used disproportionately privilege the most popular items, disregarding the long tail of less popular items—and particularly niche demands—in their recommendations.

Calling on this research gap, the contributions of our work are threefold:

1. The first main contribution relates to the quantitative measurement of a user's music mainstreaminess, which extends our previous work [56, 60]. Assuming that there is a difference between a global mainstream and a country-specific one, we quantify mainstreaminess on a global and a country level. We streamlined our mainstreaminess framework by providing distribution- and rank-based mainstreaminess measures on *artist playcounts* (APC) as well as on *artist listener counts* (ALC), resulting in 6 different measures of mainstreaminess.

2. The second main contribution of our work relates to country-specific differences in music listening behavior with respect to the degree of deviation from the global mainstream. We conducted in-depth quantitative and qualitative studies of music mainstream as evidenced in user-generated listening data from the music platform Last.fm, based on 53,259 users from 47 countries. Our results indicate that there are substantial country-specific differences with respect to the most popular artists listened to in each country. When ordering the countries according to their mainstreminess level, the order of countries depends on the underlying mainstreaminess definition. Delving into detail, we could identify three groups of countries: (i) those countries where users' music consumption behavior corresponds to the global mainstream (e.g., the United Kingdom (UK), the United States (US), the Netherlands (NL)), (ii) those countries that show a distinct country-specific mainstream that is listened to in addition to the global mainstream (e.g., Finland (FI), Brazil (BR), Russia (RU)), and (iii) those countries where the global mainstream is important in the country but, still, users listen very frequently to some artists that are not part of the global mainstream (e.g., Japan (JP), China (CN), Indonesia (IN)). Furthermore, adopting two different approaches, we identified and discussed artists with substantially higher and lower country-specific popularity (i.e., positive and negative artist outliers) for selected countries. In a qualitative analysis of outliers, we found that outliers of the same type (positive or negative) in a country commonly share genre or music style (e.g., metal bands are negative outliers in the United Kingdom (UK), but positive ones in Finland (FI)). An in-depth mixed-methods analysis of artist outliers could be an interesting future avenue of research that, on the one hand, will contribute to characterize and group countries with respect to music taste, and, on the other hand, will help predict music preferences of users with specialized, but still country-aligned, music preferences. Based on the combined results of all analyses, we conjecture that the strength and also the type of outliers (e.g., Mozart is the only classical representative in our outlier analysis and only shows up in Japan (JP)) will deliver fruitful insights for the measurement of music preferences, the characterization of countries in terms of music taste, and, in turn, build a good basis for further improving personalized music recommendations.

3. This leads to the third main contribution of our work: We demonstrated how considering a user's country in the personalized music recommendation process can notably improve accuracy of rating prediction, compared to a one-fits-all solution without country information. In doing so, we compared the performance of tailoring music recommendations to three different mainstreaminess levels (*low*, *mid*, and *high*) for each of the presented mainstreaminess definitions. This allowed us to study how the combination of user filtering with respect to mainstreaminess and to country influences the quality of music recommendations. Not surprisingly, RMSE results generally differ between users in the different mainstreaminess segments: *high*, *mid*, and *low*. With current approaches to music recommendations, it is typically difficult to satisfy listeners that have specialized music preferences far away from the global music mainstream. Our results, in contrast, show that the *low* mainstreaminess user segment is well served with our approaches. The results suggest that the combination of APC to quantify artist preferences and the rank-based mainstreaminess definition considerably outperforms the baseline (including all users in the dataset irrespective of mainstreaminess level), even for *low* mainstreaminess users; which is particularly true for the country-specific definition of mainstream.

Concluding, all measurement approaches substantially outperformed the baseline for each user set. And each of the presented mainstreaminess definitions has its particular merits. The detailed results of our study allow to devise specific user models encoding mainstreaminess and demographic information. When integrating these models into a collaborative-filtering music recommender system (MRS), the proper combination of mainstreaminess measure and scope (global or country) for a given user improves recommendation accuracy.

In this regard, a logic next step is to develop advancements in mainstreaminess measurement, which may further improve recommendation performance and serve different mainstreaminess levels equally well and (overall) superior. Another avenue of research focuses on algorithmic advancements. In our work, we could identify and demonstrate for which user group—ranging from users strongly inclined to mainstreaminess (*high* user set) to users with specialized music preferences (*low* user set)—which out of the 6 proposed mainstreaminess measures works best. This finding can be readily implemented in a recommender system: First, using the proposed measurements, the target user's mainstreaminess levels have to be computed and, considering the other users' mainstreaminess levels, the target user has to be assigned to the corresponding user set (i.e., *high*, *mid*, *low*) for each of the mainstreaminess approaches. Second, the recommendation approach that perform best for the (mainstreaminess measure, user set) pair has to be identified and applied. Taking an even more fine-tuned approach, depending on a user's country and the corresponding country profile, the corresponding most appropriate measure and/or algorithm may be adopted for the recommendation task. This approach could be summarized as "recommender of recommenders": Depending on the user profile—which gives information on the user's country and his or her *low*/*mid*/*high*-classification—the corresponding measure and approach may be selected by the system that most probably works best based on that user profile. In other words, instead of using one recommendation approach for all users equally, the most appropriate recommendation approach is selected for the given user setting, which is then used to compute the items recommended to that particular user.

Besides the already mentioned suggestions for further research, we will delve into detail and study the countries separately. Specifically, we will analyze for which country archetype (Section *Differences in music preferences related to globally popular artists*) which kind of mainstreaminess definition performs particularly well or poorly. This path of research is also needed for advancements in the above-mentioned "recommender of recommenders" approach.

In particular, we would like to encourage research endeavors that can provide explanations of causality why certain measures (e.g., distributed-based, rank-based) performed better on certain user groups (e.g., *high*, *mid*, *low*). For instance, certain user groups may be less connected with other users, thus developing music preferences that are not well served with a collaborative filtering approach. Analyzing connectedness with users within a country or with users from abroad could be a fruitful research path that could contribute to answering why country-specific or global measures, respectively, work better for some user groups than for others (e.g., [102]). Other reasons for poor performance of a particular measure for a particular user group could be rooted in the distinctiveness of the user taste. For instance, users that have developed a specialized music taste have typically a low degree of mainstreaminess; still, some users from the *low* user segment may have rather narrow music preferences (e.g., listening to merely a few artists from the long tail of popularity), whereas others' may show more variability in their listening behavior (e.g., listening to numerous different artists from the long tail of popularity). Overall, research endeavors with respect to causality would contribute fundamentally to future work on developing improved measures and to further advance recommendation algorithms.

Furthermore, in this work we treated user country as a proxy for national culture. Future work should expand the perspective on cultural aspects. For instance, differences between inhabitants of metropolitan or rural areas could show different music preference profiles. Thereby, profiles may be generalized to a global perspective or may vary within a country. In addition, other aspects that shape culture (e.g., religion or language) may be analyzed for their role in MRS.

As with every experimental study based on samples drawn from a wider population, also our study can naturally not completely eliminate the possibility that latent, confounding features potentially exist. Given the large dataset and rigorous evaluation, it is, though, unlikely that the developed features would work well by chance.

Our findings presented in this article have direct practical implications. Our presented approaches can be readily adopted in real-world MRS such as in the music streaming services Spotify or Pandora, and likewise may also serve multimedia platforms hosting music videos, such as YouTube. From a theoretical perspective, our work shows the existence of national boundaries on the global music listening platform Last.fm. This is particularly interesting, because compared to locally oriented markets (e.g., the market for food products [103]), the music recording industry is generally considered a globally oriented market [104]. While one could expect an intensification of the global orientation on the online music market, our work identified national boundaries of music listening behavior as presented on Last.fm. On top of that, incorporating the cultural aspect (by considering country specificities) of music mainstreaminess for music recommendation, our approach is a first step to promote more local initiatives. This may additionally contribute to furthering the convergence towards a less centralized and "averaged" musical taste of users, that might be hidden within the global mainstream level and also the national level.

## Acknowledgments

The authors would like to thank the Competence Center for Empirical Research Methods at WU Vienna for their support in the statistical analysis.

## Author Contributions

**Conceptualization:** Christine Bauer, Markus Schedl.

**Data curation:** Markus Schedl.

**Formal analysis:** Christine Bauer, Markus Schedl.

**Funding acquisition:** Christine Bauer.

**Investigation:** Christine Bauer, Markus Schedl.

**Methodology:** Markus Schedl.

**Project administration:** Christine Bauer.

**Validation:** Christine Bauer.

**Visualization:** Markus Schedl.

**Writing – original draft:** Christine Bauer, Markus Schedl.

**Writing – review & editing:** Christine Bauer, Markus Schedl.